\documentclass[aps,twocolumn,prd,superscriptaddress]{revtex4-1}

\usepackage[utf8]{inputenc}

\usepackage{mathtools}
\usepackage{amsfonts}
\usepackage{mathrsfs}
\usepackage{bbm}
\usepackage{slashed}

\usepackage{graphicx}
\usepackage[dvipsnames]{xcolor}
\usepackage{array}

\usepackage{placeins}

\usepackage{xspace}
\usepackage{hyperref}
\usepackage[nameinlink]{cleveref}
\usepackage{bookmark}

\usepackage{xifthen}
\usepackage{xcolor}
\usepackage{enumitem}
\usepackage{booktabs}
\usepackage{units}
\usepackage{comment}
\hypersetup{
	colorlinks,
	linkcolor={red!75!black},
	citecolor={blue!75!black},
	urlcolor={blue!75!black}
}
\usepackage{units}

\usepackage[utf8]{inputenc}

\setkeys{Gin}{width=0.48\textwidth}

\newcommand{\imag}{\text{i}}

\def\0#1#2{\frac{#1}{#2}}

\def\bea{\begin{eqnarray}}
\def\eea{\end{eqnarray}}
\def\be{\begin{equation}}
\def\ee{\end{equation}}

\newcommand{\Tr}{\mathrm{tr}}

\newcommand{\pToFigs}{.}

\def\CP{{\mathcal P}}

\newcommand{\CPC}{{\langle \varphi \varphi^\dagger\rangle_{\rm c}}}
\newcommand{\PP}{\tilde P}
\newcommand{\pt}{{\bf p}}
\newcommand{\xt}{{\bf x}}
\newcommand{\tcond}{t_{\text{cond}}}
\newcommand{\Q}{Q}


\newcommand{\gettitle}{Order parameters for gauge invariant condensation far from equilibrium}

\hypersetup{
	colorlinks,
	linkcolor={red!75!black},
	citecolor={blue!75!black},
	urlcolor={blue!75!black},
	pdftitle={\gettitle},
	pdfauthor={Berges, Boguslavski,Pawlowski},
	pdfkeywords={Spectral function} {Non-thermal fixed points}
	{Order parameters} {Gluon} 
	{Scaling}
	{Real time} {Infrared},
	bookmarksopen=true,
	bookmarksopenlevel=2,
	bookmarksnumbered=true
}

\begin{document}

\title{\gettitle}

\author{Jürgen Berges} 
\affiliation{Institute for Theoretical Physics, 
  Heidelberg University, Philosophenweg 12, 69120 Germany}

\author{Kirill Boguslavski} 
\affiliation{Institute for Theoretical
  Physics, Technische Universit\"{a}t Wien, 1040 Vienna, Austria}

\author{Lillian de Bruin}\affiliation{Institute for Theoretical Physics, 
  Heidelberg University, Philosophenweg 12, 69120 Germany}

\author{Tara Butler}\affiliation{Institute for Theoretical Physics, 
  Heidelberg University, Philosophenweg 12, 69120 Germany}\affiliation{LIX, École Polytechnique, CNRS, IP Paris,
91120 Palaiseau, France}

\author{Jan~M.~Pawlowski} 
\affiliation{Institute for Theoretical Physics, 
  Heidelberg University, Philosophenweg 12, 69120 Germany}

\begin{abstract}
Nuclear collisions at sufficiently high energies are expected to produce far-from-equilibrium matter with a high density of gluons at early times. We show gauge condensation, which occurs as a consequence of the large density of gluons. To identify this condensation phenomenon, we construct two local gauge-invariant observables that carry the macroscopic zero mode of the gauge condensate. The first order parameter for gauge condensation investigated here is the correlator of the spatial Polyakov loop. We also consider, for the first time, the correlator of the gauge invariant scalar field, associated to the exponent of the Polyakov loop. Using real-time lattice simulations of classical-statistical $SU(2)$ gauge theory, we find gauge condensation on a system-size dependent time scale $t_{\text{cond}} \sim L^{1/\zeta}$ with a universal scaling exponent $\zeta$. 
Furthermore, we suggest an effective theory formulation describing the dynamics using one of the order parameters identified. The formation of a condensate at early times may have intriguing implications for the early stages in heavy ion collisions. 
\end{abstract}

\maketitle

\section{Introduction}
Understanding the properties of matter in extreme conditions is a fundamental pursuit in the field of high-energy physics. 
In particular, the study of the quark-gluon plasma (QGP) formed in high-energy collision experiments involving heavy nuclei has provided valuable insights into the fundamental nature of strong interactions governed by quantum chromodynamics (QCD) \cite{Gelis:2010nm, Lappi:2006fp}. 

In recent years, there has been growing interest in the possibility of a condensate emerging during the early time evolution of the QGP. 
The matter formed moments after a heavy ion collision is far from equilibrium and is characterized by a large initial density of gluons, which could facilitate the formation of a condensate \cite{Blaizot:2011xf}. 
However, this picture has several complications.
First, classical-statistical and kinetic theory simulations have shown that the plasma evolution does not support the formation of a Bose condensate of gluon fields \cite{Kurkela:2012hp, Berges:2013eia, AbraaoYork:2014hbk, Blaizot:2016iir}. 
Complications also arise due to the non-Abelian nature of the gauge theory: for the direct identification of gauge condensation one must construct a \textit{local} gauge-invariant operator which  measures the macroscopic zero mode that signals gauge condensation. Previous examples of gauge-invariant operators for nonequilibrium condensation have been studied in the context of the Abelian Higgs model and its relation to non-Abelian theories \cite{Gasenzer:2013era, Ford:1998bt, Mitreuter:1996ze}. 

It has been demonstrated in \cite{Berges:2019oun}, that the initial overoccupation of gluons does in fact lead to the formation of a gauge-invariant condensate. Condensation is tantamount to the formation of a macroscopic zero-mode expectation value that scales proportionally with $(2\pi)^d\delta^{(d)}(0)\rightarrow L^d$ for a $d-$dimensional finite volume with the length scale $L$ \cite{Berges:2012us}. Specifically, we studied the expectation value of the spatial Wilson loop, which computes the infrared excitations of non-Abelian gauge theories \cite{Berges:2007re, Dumitru:2014nka, Mace:2016svc, Berges:2017igc}. A volume-independent condensate fraction was established for increasingly $L\rightarrow \infty$. It was further demonstrated that condensate formation is a far-from-equilibrium phenomenon, while the zero mode fraction does not grow in time or scale with volume in (classical) thermal equilibrium as expected.
However, 
the Wilson loop is a non-local object. Therefore it cannot be used to formulate a low energy effective theory with which we can further unravel the dynamics of the infrared condensate. 

In this paper, we introduce two local order parameters that are related to the spatial Wilson loop: the first one is the two-point connected correlation functions of the traced spatial Polyakov loop. The second one is based on the algebra-valued gauge-invariant scalar field, that can be constructed from the exponent of the Polyakov loop. These fields are gauge invariant and live in a $2+1$ dimensional space-time. We will demonstrate that their correlators capture the same condensation phenomena as the Wilson loop while being sufficiently local objects.

In particular, by employing classical-statistical lattice simulations, we investigate the dynamics of the Yang-Mills plasma with large gluon densities and weak coupling. 
The simulated system approaches a self-similar regime with universal scaling properties, which has been studied in detail \cite{Berges:2013eia,Berges:2013fga,Berges:2017igc} and shows similarities with scalar systems \cite{Berges:2014bba, Berges:2015ixa}. 
We extract the condensation observable in the vicinity of this self-similar state and analyze its evolution over time. 
Through these investigations, we aim to shed light on the nature of gauge condensates, their implications in hydrodynamics and transport phenomena, and their potential role in the formulation of a low-energy effective field theory for QCD. By studying local correlation functions of an algebra-valued scalar field, we can compare our results with far-from-equilibrium Bose condensation for a scalar order parameter field \cite{Berges:2012us, Orioli:2015dxa, Schachner:2016frd, Walz:2017ffj, Chantesana:2018qsb, Boguslavski:2019ecc}.

This paper is organized as follows. In \Cref{sec:setup} we first discuss the early-time infrared dynamics in the far-from-equilibrium QGP and then introduce the local order parameters used to identify condensation. Then, in \Cref{sec:condensation} we discuss our classical-statistical lattice simulations and present results that demonstrate condensate formation. We finally conclude in \Cref{sec:conclusion}.

\begin{figure}[tp!]
\begin{center}
\includegraphics[width=0.75\linewidth]{\pToFigs/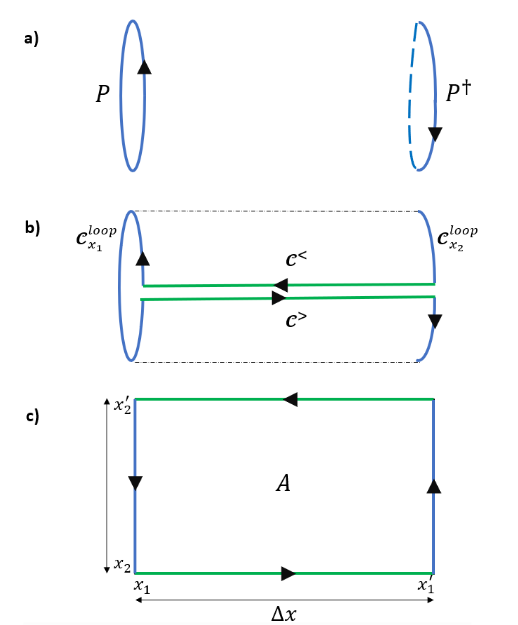}
\caption{Visualization of the spatial correlation between the Polyakov loop correlator $\langle PP^\dagger \rangle (t, \Delta x, L)$ in a) and the rectangular spatial Wilson loop $W[t,\Delta x, L]$ in c). The spatial periodicity of the system results in the identification of coordinates $x_2' \equiv x_2\ \textrm{mod}\ L$ on the lattice, causing the Wilson loop to exhibit a wrapping effect in the $x_2$ direction, as depicted in b). The distinguishing feature between the Polyakov loop correlator and the Wilson loop is revealed to be the presence of connecting Wilson lines (highlighted in green) in the latter, demonstrating the connection between the two quantities.
}
\label{fig:DepictionPol_Wil}
\end{center}
\end{figure}
%

\section{Early-time infrared dynamics far from equilibrium}\label{sec:setup}

In Ref.~\cite{Berges:2019oun} it has been shown that the early time evolution of QCD exhibits a condensate. 
The respective signatures have been obtained from the scaling dynamics of (large) spatial Wilson loops. While this analysis demonstrates the presence of gauge condensates, their nature and dynamics remain to be unravelled, which we address in this paper.

This is of great importance for the formulation of non-Abelian gauge theories, specifically for describing the early time dynamics of QCD in terms of an effective kinetic theory that stems from an effective Lagrangian. In the present Section we briefly discuss a corresponding reparameterization of the Yang-Mills action, 
\begin{equation}
\begin{split}
S_\textrm{YM}[A] =&\, - \frac12 \int_x \Tr \,F_{\mu\nu} F^{\mu\nu}\,,\\[1ex]
 F_{\mu\nu}=&\, \partial_\mu A_\nu -\partial_\nu A_\mu-\imag g[A_\mu,A_\nu]\,,
\end{split}
\label{eq:SYM}
\end{equation}
with the trace in the fundamental representation. We formulate the theory in a manifestly infrared finite setup with periodic boundary conditions on a torus $T^3$,
\begin{align}
\int_x = \int \textrm{d} t \int_{T^3} \textrm{d}^3 x\,,\qquad T^3 =\mathbbm{R}^3\ \textrm{mod}\ L \,.
\end{align}
We have put the Yang-Mills theory in a finite spatial box for several reasons. To begin with, it reflects the situation in a heavy ion collision with a rapidly expanding but finite fireball. Moreover, it allows for a gauge-invariant regularization and control of infrared divergences. 
Finally, in the present work we employ simulations to compute the far-from-equilibrium dynamics of gauge theories that are formulated on a finite spatial lattice with periodic boundary conditions.

\Cref{eq:SYM} is formulated using gauge fields, which are gauge variant degrees of freedom. However, due to the presence of gauge condensation, it is more desirable to rewrite the Yang-Mills action partially in terms of gauge-invariant degrees of freedom that may directly carry the condensation phenomenon.
The expectation value of spatial rectangular Wilson loops, 
\begin{align}
W[t,\Delta x,L] = \frac{1}{N_c} \Tr \, \CP  e^{-i\, g \int_{\cal C} A_i(t,\boldsymbol{x})\, d x_i}\,,
\label{eq:SpatialWilson}
\end{align}
serves as a gauge-invariant order parameter for gauge condensation \cite{Berges:2019oun}, where $\CP$ denotes path ordering and ${\cal C}$ is the closed rectangular path. In \labelcref{eq:SpatialWilson}, the Wilson loop stretches over the whole $x_2$-direction of length $L$ and has an extent $\Delta x=x_1-x_1'$ in $x_1$ direction, as depicted in the lower panel of \Cref{fig:DepictionPol_Wil}. Accordingly, it is a non-local observable instead of a correlation function of local dynamical degrees of freedom. In \Cref{sec:local_order_params} we will construct local observables in the $x_1$ direction that are derived from the Polyakov loop, whose correlator is depicted in the upper panel of \Cref{fig:DepictionPol_Wil}.

\subsection{Hierarchy of scales far from equilibrium}

In thermal equilibrium at high temperature $T$, the weakly-coupled QCD plasma exhibits a hierarchy of scales: 
hard momenta $\sim T$ that dominate the energy density of the system, a soft electric screening (Debye) scale $\sim gT$, and an ultrasoft magnetic screening scale $\sim g^2T$, where $g^2=4\pi\alpha_s\ll 1$. 

Far from equilibrium, such a separation of scales also exists in the vicinity of the non-thermal attractor, the self-similar scaling regime of highly occupied plasmas that is insensitive to the initial conditions and the exact value of the coupling \cite{Kurkela:2012hp,AbraaoYork:2014hbk,Schlichting:2012es, Berges:2017igc,Berges:2013eia,Berges:2013fga,Berges:2008mr,Kurkela:2011ti,Boguslavski:2019fsb}. The hard scale, which dominates the system's energy density, scales as $\Lambda(t)\sim t^{1/7}$, the soft electric screening (Debye) mass follows $m_D(t)\sim t^{-1/7}$, and the ultrasoft magnetic screening scale evolves as $\sqrt{\sigma}(t)\sim t^{-\zeta/2}$, which is associated with the string tension $\sigma$ and with the sphaleron transition rate \cite{Kurkela:2011ti, Kurkela:2012hp, Berges:2017igc, Berges:2013fga, Mace:2016svc, Moore:1997cr}. These characteristic momentum scales are initially of the same order $Q_s$, but the self-similar evolution leads to a dynamical separation of scales over time, such that $\sqrt{\sigma}(t)\ll m_D(t)\ll \Lambda(t)$. 
The ultrasoft scale approaches zero faster than the Debye screening scale in time, and its scaling exponent sets the time scale for the build-up of a coherent macroscopic state--our condensate. 

Below the magnetic scale, at ultrasoft momenta where the occupation numbers are expected to be $f\sim 1/\alpha_s$, the dynamics is non-perturbative. In this regime, the notion of gauge-fixed particle numbers based on a distribution $f$ of gauge modes is ill-defined. Therefore, we cannot approach condensation by counting occupancies of quasi-particle states. Furthermore, far from equilibrium, a condensate can emerge without having a well-defined chemical potential entering its distribution \cite{Berges:2012us, Orioli:2015dxa}, in contrast to condensation in thermal equilibrium. 

We employ a more general approach to condensation, as this phenomenon can be identified from properties of correlation functions in strongly correlated systems, both in and out of equilibrium. The previous study \cite{Berges:2019oun} investigated the dynamics at long distances using gauge-invariant but non-local spatial Wilson loops. The extension of this study, in order to eventually formulate a kinetic theory describing this phenomenon, necessitates the investigation of more local order parameters.

\subsection{Local order parameters}
\label{sec:local_order_params}

\begin{figure}[t]
\begin{center}
\includegraphics[width=1.0\linewidth]{\pToFigs/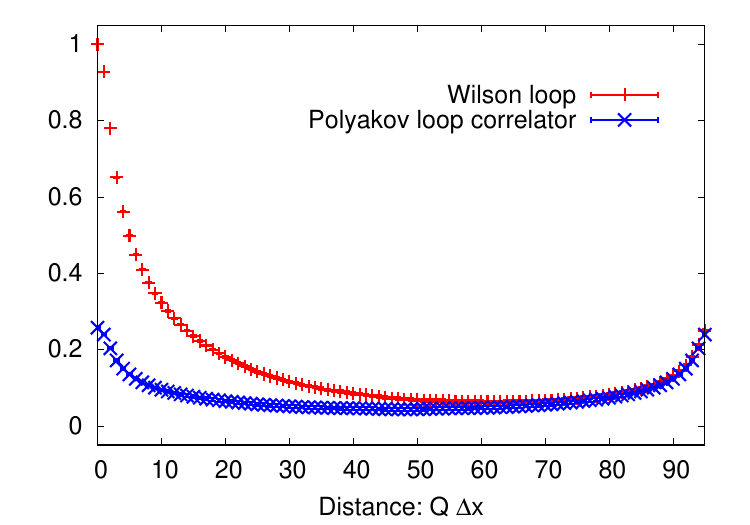}
\caption{Comparison of the Wilson loop expectation value $\langle W\rangle$ and the Polyakov loop correlator $\langle PP^{\dag}\rangle$ on a lattice with size $N_s=96$ at time $Qt=1000$. The Polyakov loop correlator $\langle PP^{\dag}\rangle$ exhibits consistent dynamics with the Wilson loop at larger spatial separations, indicating a strong correlation between the two quantities.}
\label{fig:Pol_Wil_illustr}
\end{center}
\end{figure}

We seek for degrees of freedom that do not hide essential parts of the dynamics by absorbing them, as the Wilson loop \labelcref{eq:SpatialWilson} does, but are as local as possible. The latter is important for the construction of a (local) kinetic theory for QCD,  specifically to rewrite the action \labelcref{eq:SYM} in terms of these degrees of freedom. 

This becomes clear when we view the Wilson loop as a correlation function instead of a dynamical field. Let us consider the Wilson loop in \labelcref{eq:SpatialWilson}, shown in the lower panel of \Cref{fig:DepictionPol_Wil}, that winds around the full extent of the $x_2$ direction. This observable gives us access to the infrared dynamics that we are interested in. From the simulation point of view, it also smooths out ultraviolet effects and hence lattice artifacts %
\footnote{We note that in classical-statistical simulations of highly occupied systems far from equilibrium the dependence on the ultraviolet modes can be further reduced if the characteristic dynamical scale $\sim Q_s$ is much lower.}.
Therefore, the Wilson loop can be understood as the correlation function of an operator at $x_1$ and one at $x_1'$. However, further non-localities are introduced by the connecting lines (green lines in the middle panel of \Cref{fig:DepictionPol_Wil}). 

This motivates the construction of a more local version of the Wilson loop that carries the correlation structure of the latter. To that end we introduce the following spatial Wilson loop, 
\begin{align}
  \label{eq:PolyakovLoopTr}
  P_i(t,\boldsymbol{x}) &=\0{1}{N_c} \Tr \, {\PP}_i(t,\boldsymbol{x}) \\
  {\PP}_i(t,\boldsymbol{x}) &= \CP  e^{-i\, g \int_{0}^{L} A_i(t,\boldsymbol{x})\, d x_i}
  \,,
  \label{eq:PolyakovLoop}
\end{align}
which corresponds to a closed path over the full $x_i$ direction, as depicted in the upper panel of \Cref{fig:DepictionPol_Wil}. It can be understood as the spatial version of the Polyakov loop that usually corresponds to a temporal Wilson loop and is used in thermal QCD as an order parameter for the confinement-deconfinement phase transition. We shall therefore call it the spatial Polyakov loop, or in short, the Polyakov loop in a slight abuse of notation. Its two-point function with $\boldsymbol{x},\boldsymbol{x}'$ differing by $\Delta x= x_1-x_1'$
\begin{align}
\langle P_2(t,\boldsymbol{x}) P^\dagger_2(t,\boldsymbol{x}')\rangle 
\label{eq:PolTwoPoint}
\end{align}
is related to the expectation value of the Wilson loop \labelcref{eq:SpatialWilson} as suggested by \Cref{fig:DepictionPol_Wil} but lacks the non-local connection lines between $x_1$ and $x_1'$. 
Therefore, in contrast to the Wilson loop in \labelcref{eq:SpatialWilson}, the correlation function \labelcref{eq:PolTwoPoint} is symmetric under $\Delta x \to L - \Delta x$ due to spatial periodicity. This can be seen in \Cref{fig:Pol_Wil_illustr} where both quantities are plotted as functions of the distance $\Delta x$. While for small distances the non-local connection lines seem to be important, their impact decreases at larger distances where both quantities agree.

Still, \labelcref{eq:PolyakovLoopTr} wraps around the $x_i$-direction and one may ask whether this apparent non-locality can be removed. To begin with, while we are interested in the infrared dynamics of QCD, a local operator inevitably also carries ultraviolet fluctuations that have to be regularized. Moreover, the non-local gauge symmetry prohibits the construction of fully local operators that do not carry (part) of the dynamics as discussed in the beginning. In short, \labelcref{eq:PolyakovLoopTr} is a gauge-invariant quantity that is sensitive to the infrared but sufficiently local in space.

In the remainder of this Section, we introduce a scalar field via the (traced) spatial Polyakov loop \labelcref{eq:PolyakovLoopTr} and discuss how it emerges naturally as a variable in the action \labelcref{eq:SYM}. 
To that end, we note that the operator in the trace \labelcref{eq:PolyakovLoop} can be written as 
\begin{align}
 {\PP}_i(x) = \CP  e^{-i\, g \int_{0}^{L} \mathcal{A}_i(x)\, d x_i} = e^{\imag \phi_i(x)}\,,
 \label{eq:PolyakovLine}
\end{align}
where we use the same notation $x = (t,\boldsymbol{x})$ as before. 
Here we have introduced the algebra-valued field 
\begin{align}
\phi_i(x)=\phi_i^a(x)\, t^a\,, \qquad t^a =\sigma^a/2\,,
\label{eq:phia}
\end{align}
with the Pauli matrices $\sigma^a$, $a=1,2,3$, for $\PP_i \in $ SU(2). This invites us to define the gauge-invariant scalar field $\varphi_i(x)$ as
\begin{align}
    \0{1}{N_c} \Tr \PP_i \equiv P_i = \cos \varphi_i.
\end{align}
Just as $P_i$, the scalar field $\varphi_i(x)$ lacks any dependence on the spatial direction $x_i$ that has been integrated out. 
On the other hand, the spatial Polyakov lines transform covariantly under gauge transformations $U\in$ SU(2), with ${\PP}_i\to U {\PP}_i U^\dagger$. It follows that the algebra element transforms in the same way,
\begin{align}
\phi_i(x)\to U(x) \phi_i(x) U^\dagger(x)\,.
\end{align}
The gauge transformation can be used to diagonalize $\PP_i$ and hence $\phi_i$. This fixes the gauge freedom to a (spatial) Polyakov or diagonalization gauge, where the (eigenvalue) field $\varphi_i$ becomes proportional to the algebra field
\begin{align}
\phi_i(x)=\varphi_i(x) t^3\,.
\label{eq:varphi}
\end{align}
We emphasize that the scalar field $\varphi_i$ is gauge invariant since it is defined as an eigenvalue of the operator $\phi_i$. 
Moreover, $\varphi_i$ is a suitable infrared degree of freedom that we have searched for and is directly linked to the gauge fields $A_i(x)$. To see this, we consistently use the Polyakov gauge for $\PP_i$, $\phi_i$, and for $A_i(x)$ at all $x_i$, which corresponds to rotating them into the Cartan subalgebra. Then the path ordering in \labelcref{eq:PolyakovLine} can be dropped and we have the relation 
\begin{align}
\phi_i(x) = - g \int_0^L \textrm{d} x_i A_i(x) =- g L A_i(x) \,, 
\label{eq:phi-A}
\end{align}
where the last step can be done if the gauge transformation is also used to remove the $x_i$ dependence in the gauge field $A_i(x)$. In the temporal direction such a gauge transformation enforces the Polyakov gauge. Note that in the evaluation above this is not introduced as a gauge but for elucidating the relation between the phase fields $\phi_i$ and the gauge field. 
However, this also entails that we can use a respective gauge in \labelcref{eq:SYM} to make the dependence of the action on the gauge covariant phase field $\phi_i$ and hence on the gauge invariant field $\varphi_i$ apparent. Since we only want to illuminate this connection, we use this gauge in \labelcref{eq:SYM} in the following; the respective gauge fixing has been discussed in detail in \cite{Gasenzer:2013era} and literature therein. 

Let us now single out the spatial direction $x_3$, choose the spatial Polyakov gauge as before, and reformulate the classical action in \labelcref{eq:SYM} in terms of $\phi_3(x)$. 
To that end we invert the relation  \labelcref{eq:phi-A} and substitute in \labelcref{eq:SYM}
\begin{align}
A_3(x) = - \frac{1}{g L} \phi_3(x) =- \frac{1}{g L} \varphi_3(x) t^3\,.
\end{align}
This leads us to the formulation of the classical action
\begin{subequations}
\label{eq:SYMphi}
\begin{align}
S_{\textrm{YM}}=\frac12 \int_x \Tr\, F_{\bar\mu \bar \nu}^2 +\int_x \Tr \,F_{3\bar\mu}^2 \,, 
\label{eq:SYM-3barmu}
\end{align}
where $\bar \mu=0,1,2$ and $\bar x=(x_0,x_1,x_2)$, and 
\begin{align}\nonumber 
\int_{x} \,\Tr F_{3\bar\mu}^2 = &\,\frac12\int_x \Tr \,(\partial_3 A_{\bar \mu})^2+ \frac{g}{L} \int_{\bar x} \Tr\,\phi_3(\bar x)\int_{x_3} [A_{\bar \mu}, \partial_3 A_{\bar \mu}]\\[2ex] 
& + \frac{1}{g^2  L^2}\int_x \Tr (D_{\bar \mu}\phi_3)^2\,, 
\label{eq:Non-AbelianHiggs}
\end{align}
\end{subequations}
with the covariant derivative $D_\mu =\partial_\mu- \imag g A_\mu$. 
Here the gauge-invariant degree of freedom $\varphi_3$ enters directly.

Since we aim at establishing a kinetic theory formulation for the scalar field eventually, we make an additional approximate simplification. For the moment, we neglect $\partial_3 A_{\bar \mu}$ since it is expected to have little impact on the dynamics of $\phi_3$. 
Then the first line of \labelcref{eq:Non-AbelianHiggs} vanishes, and the second line, together with the pure gauge field part \labelcref{eq:SYM-3barmu}, comprises the action of a non-Abelian Higgs model in 2+1 dimensions, with
\begin{align}
S_{\textrm{YM}}\to L\left[\frac{1}{2} \int_{\bar x} \Tr\, F_{\bar\mu \bar \nu}^2 + \frac{1}{g^2  L^2}\int_x \Tr (D_{\bar \mu}\phi_3)^2\right]\,. 
\end{align}
However, with or without this last step, \labelcref{eq:SYMphi} illuminates the dependence of the Yang-Mills action on the algebra field $\phi_3$. The latter, in a conveniently chosen gauge, carries the dynamics of the respective gauge field component $A_3$. It is also suggestive that this allows for the construction of effective actions in terms of the algebra fields, well- or naturally suited to describe the low-energy equilibrium and infrared far-from-equilibrium dynamics of gauge theories. 

In the present work, we study the dynamics of the gauge condensate in terms of the two-point correlations of both the gauge-invariant Polyakov loop $P$ and the gauge-invariant field $\varphi$ defined via \labelcref{eq:PolyakovLoopTr} and \labelcref{eq:varphi}. Here and in the following we will simplify the notation by setting $P \equiv P_2$ and $\varphi \equiv \varphi_2$. 
In a forthcoming study we also investigate their higher order fluctuation observables, $\langle \varphi(t_1, \boldsymbol{x}_1) \cdots \varphi(t_n, \boldsymbol{x}_n)\rangle$ as a further important step towards the formulation of an effective theory of QCD in terms of these algebra fields. 

\begin{figure}
    \centering
    \includegraphics[width=1.0\linewidth]{\pToFigs/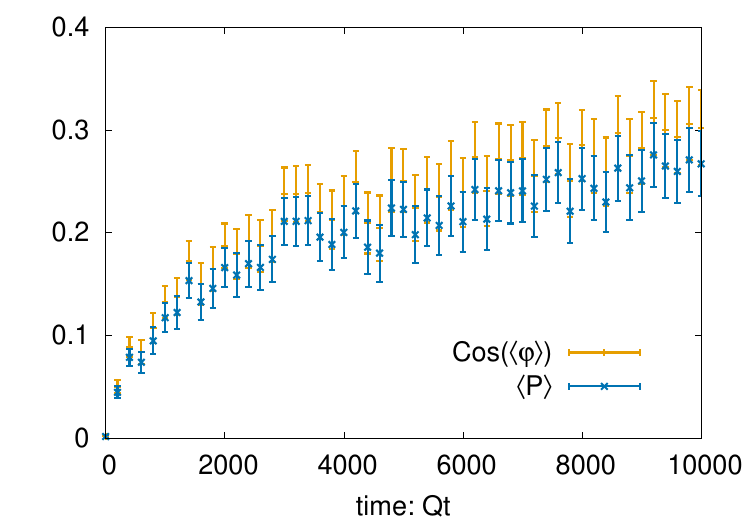}
    \caption{Comparison of the expectation value of the Polyakov loop to the Polyakov loop re-expressed in terms of the algebra-valued holonomous eigenvalue field for $N_s=96$ lattice, $Q_st=1000$. This shows the approximate agreement of the one-point functions $\langle P \rangle \equiv \langle \cos (\varphi) \rangle \approx \cos (\langle \varphi \rangle)$. 
    }
    \label{fig:PvsCosPhi}
\end{figure}
\begin{figure*}[t]
\includegraphics[width=0.47\linewidth]{\pToFigs/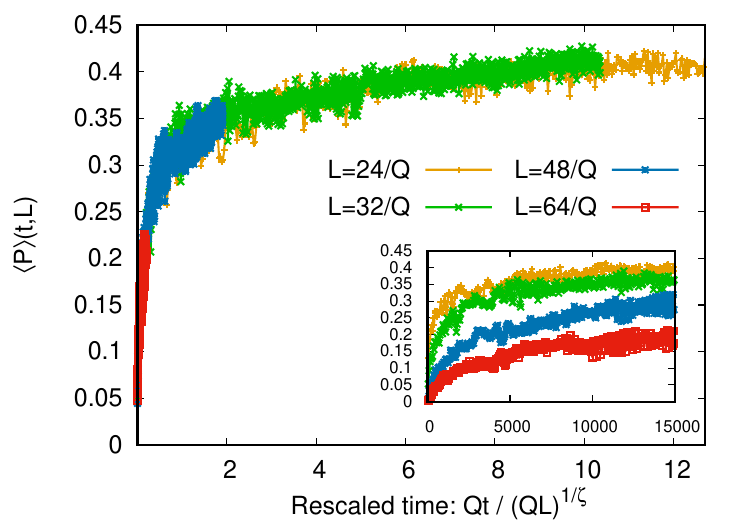}
\includegraphics[width=0.47\linewidth]{\pToFigs/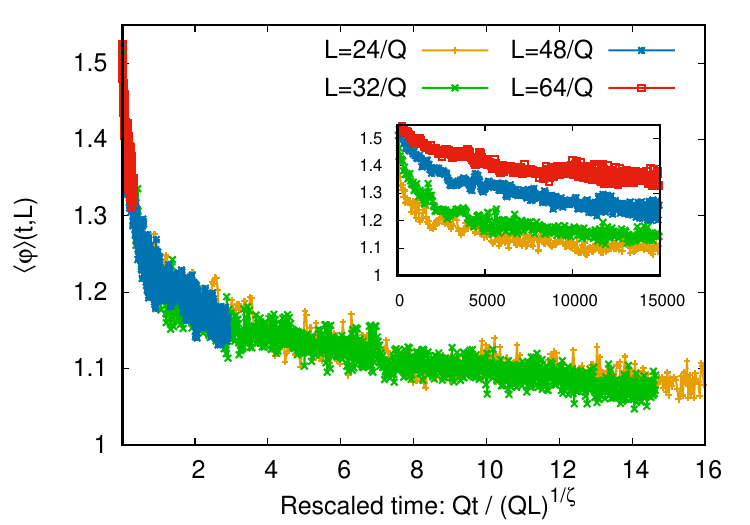}
\caption{Expectation values of the spatial Polyakov loop and the scalar algebra field as functions of rescaled time for different volumes. The Polyakov loop is rescaled with $\zeta=0.24$ and the algebra field is rescaled with $\zeta=0.25$, both exhibiting scaling behavior in time. We also observe that $\langle\phi\rangle$ starts from $\approx \pi/2$ and decreases with time, which is in line with \Cref{fig:PvsCosPhi}. In the context of condensation, the connected correlator becomes increasingly independent of $\langle\phi\rangle$, indicating Bose-Einstein condensation behavior.
}
\label{fig:Scaling_P_Phi}
\end{figure*}
%

\section{Gauge-invariant condensation}\label{sec:condensation}

With $\varphi$ and $P$ given by \labelcref{eq:PolyakovLoopTr} and \labelcref{eq:varphi}, we have introduced two local order parameters that can be employed to study the dynamics of the gauge condensate as an alternative to the non-local Wilson loop. Next, we turn our attention to studying the gauge condensate itself. 

First, we will discuss the lattice simulations employed to evaluate the non-perturbative real-time evolution. Then, we will define our observables and discuss the signals for condensation. Finally, we will discuss the condensate fraction and the time evolution of the condensate.

\subsection{Classical-statistical lattice simulations}

The initial gluons produced in high energy heavy ion collisions carry momenta on the order of the saturation scale $Q$, at time $t\sim1/Q$ in natural units \cite{Gelis:2010nm, Lappi:2006fp}. The system is considered strongly correlated despite the small running gauge coupling $\alpha_s(Q)$ due to the high initial gluon occupancies $f\sim 1/\alpha_s(Q)$. It follows that such a non-perturbative problem can be studied via classical-statistical lattice simulations \cite{Aarts:2001yn, Smit:2002yg}. 
The characteristic initial over-occupation of gluons is translated into energy density $\sim Q^4/\alpha_s$ and fluctuations, which initializes the lattice gauge theory evolution. Throughout this study, quantities are given in terms of $Q$.

We discretize the $SU(N_c)$ gauge theory with $N_c = 2$ colors on a lattice with three spatial dimensions of size $N_s$ and spacing $a_s$. The lattice gauge theory evolution is initialized as a superposition of transversely polarized gluon fields,
\begin{eqnarray}
\mathcal{A}_j^a(t=0,\pt)=\sqrt{\frac{f(0,p)}{2p}}\sum_{\lambda}c_{\pt}^a\xi_j^{(\lambda)}(\pt) + {\rm c.c.},
\end{eqnarray}
and their time derivatives,
\begin{eqnarray}
E_j^a(t=0,\pt)=\sqrt{\frac{f(0,p)p}{2}}\sum_{\lambda}c_{\pt}^a \xi_j^{(\lambda)}(\pt) + {\rm c.c.}
\end{eqnarray}
Here, $\pt$ denotes spatial momenta with $p=|\pt|$, $\xi_j^{(\lambda)}$ the transverse polarization vectors, and $c_{\pt}^a$ complex Gaussian random numbers with vanishing mean and unit variance. Index $a=1,..., N_c^2-1$ is the color index and $j=1,2,3$ is the spatial index. The initial gluon over-occupation is parameterized by (with $A = 1.14$)
\begin{eqnarray}
    f(0,p)= \frac{Q A}{4\pi\alpha_s p}\,\theta(Q-p).
\end{eqnarray}
The real-time evolution is realized by solving the classical Hamilton equations of motion in the temporal axial gauge $\mathcal{A}_0=0$. The equations of motion are formulated in a gauge covariant way, using $E_a^j(t,\xt)$ and link fields $U_j(t,\xt)=\exp(ig\alpha_a\mathcal{A}_j(t,\xt))$. For a detailed description of this standard technique we refer, e.g., to \cite{Ambjorn:1990pu, Berges:2013fga, Boguslavski:2018beu}. 

Our simulations are conducted on cubic lattices with $N_s= 48, 64, 96, 128$ lattice sites and the lattice spacing $Q a_s= 0.5$ (with $L = N_s a_s$). To obtain sufficient statistics, we average our observables over 200 configurations.

\begin{table}[t]
\begin{center}
\begin{tabular}{|c||c|c|}
 \hline
 Observable & $\Delta x$ & Exponent\\
 \hline\hline 
 $\langle PP^\dagger \rangle_c (t, \Delta x, L)$&  $L/2$ & $\zeta=0.36 \pm 0.04 $\\ 
 $\langle PP^\dagger \rangle_c (t, \Delta x, L)$&  $L/4$ & $\zeta=0.31 \pm 0.09$\\ 
 \hline
 $\CPC (t, \Delta x, L)$&  $L/2$ & $\zeta=$ 0.37 $\pm$ 0.03\\ 
 $\CPC (t, \Delta x, L)$&  $L/4$ & $\zeta=$ 0.34 $\pm$ 0.03\\
 \hline\hline 
$\langle PP^\dagger \rangle (t, \Delta x, L)$&  $L/2$ & $\zeta=0.31 \pm 0.04$\\
$\langle PP^\dagger \rangle (t, \Delta x, L)$&  $L/4$ & $\zeta=0.27 \pm 0.06$\\
\hline
$W(t, \Delta x, L)$ & $L/2$ & $\zeta=0.27 \pm 0.04$ \\
$W(t, \Delta x, L)$ & $L/4$ & $\zeta=0.24 \pm 0.06$ \\
\hline
\end{tabular} 
\caption{\label{tab:exponents-summary} Summary of scaling exponents for different condensate observables. These values and their uncertainties have been estimated using a similar $\chi^2$-procedure as in \cite{Berges:2013fga, Orioli:2015dxa, Berges:2019oun}, which we outline in \Cref{app:zeta_estimate}. The scaling exponent $\zeta$ is compared for the two sufficiently local observables studied in this work, the connected Polyakov loop correlator and the algebra field, to the Polyakov loop correlator in \labelcref{eq:full-corrs}, and to the Wilson loop expectation value studied in \cite{Berges:2019oun}. }
\end{center}
\end{table}
%

\subsection{Evolution of one-point functions}

Before defining and discussing condensates, let us first consider the evolution of the one-point functions $\langle P \rangle$ and $\langle \varphi \rangle$. Often considering the expectation values of order parameters is sufficient to describe condensation phenomena. However, in our case this would be misleading because $\langle P \rangle$ grows while $\langle \varphi \rangle$ decreases with time. 

Their relation is depicted in \Cref{fig:PvsCosPhi}, where $\langle P \rangle$ and $\cos (\langle \varphi \rangle)$ are shown as functions of time. One observes their approximate agreement
\begin{align}
    \langle P \rangle \equiv \langle \cos (\varphi) \rangle \approx \cos (\langle \varphi \rangle).
\end{align}
Since they start close to zero due to the large initial occupancies, the scalar field is initially $\langle \varphi \rangle \approx \pi/2$ and then decreases with time. This would na\"ively contradict the emergence of a condensate. 

However, one finds volume scaling in the time evolution of both $\langle P \rangle$ and $\langle \varphi \rangle$. This is shown in \Cref{fig:Scaling_P_Phi} where these one-point functions are shown for different volumes as functions of rescaled time in the main plots and original time $Q t$ in the insets. 
The rescaling proceeds by dividing time according to $Qt / (QL)^{1/\zeta}$ with the scaling exponents $\zeta=0.24$ and $\zeta=0.25$ for the Polyakov loop and the algebra field expectation values, respectively %
\footnote{More precisely, we obtain $\zeta=0.24\pm0.03$ and $\zeta=0.25\pm0.03$ by employing the method presented in \Cref{app:zeta_estimate}.}.
In this case, curves for different volumes fall on top of each other, indicating universal dynamics on a size-dependent time scale. We will argue in the following section using two-point correlators that this time scale is associated to a condensate formation time. 

Although the macroscopic field $\langle \varphi \rangle$ with time, a condensate can still emerge in the zero mode of the connected two-point function $\CPC$ that will be introduced shortly. Such a phenomenon is not very usual. For instance, scalar quartic models of inflation \cite{Micha:2002ey} can have a decaying inflaton $\langle \varphi \rangle$ while the zero mode of the connected correlator $\CPC(p{=}0)$ grows with time indicating the onset of condensation \cite{Berges:2013lsa, Orioli:2015dxa}. This serves as a motivation for us to define condensate fractions as connected correlators of the order parameters.

\subsection{Condensate fractions}

\begin{figure*}[t]
\includegraphics[width=0.47\linewidth]{\pToFigs/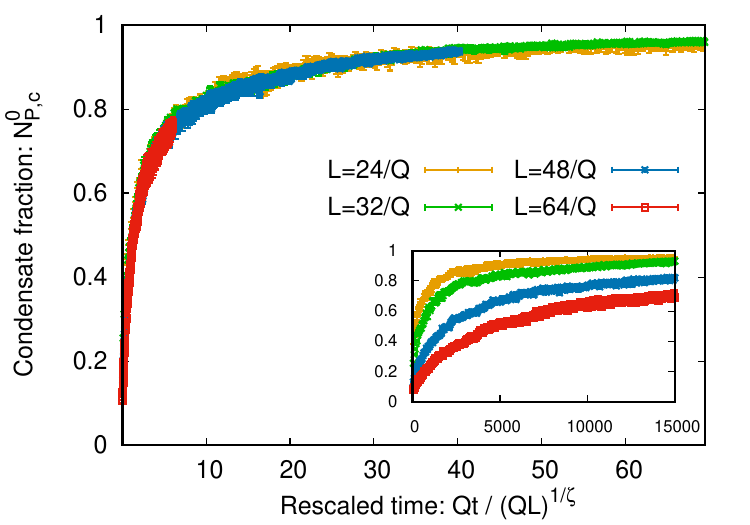}
\includegraphics[width=0.47\linewidth]{\pToFigs/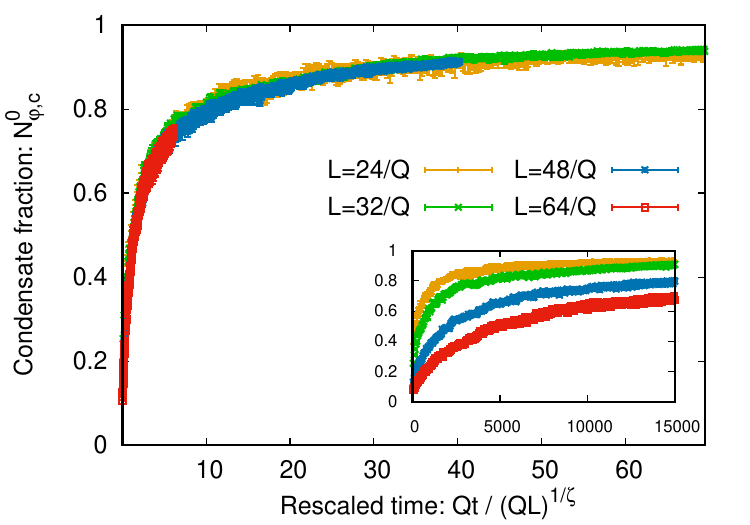}
\caption{Condensate fractions \labelcref{eq:conn-CFs} of the connected correlators $\langle PP^{\dag}\rangle_c$ and $\CPC$ as functions of rescaled time for different lattice volumes. All curves fall on top of each other and therefore show the emergence of a volume-independent condensate fraction. The rescaled quantities have scaling exponents $\zeta=0.31$ and $\zeta=0.34$, respectively, using the procedure in \Cref{app:zeta_estimate} (cf., \Cref{tab:exponents-summary}). {(\em Insets:)} Condensate fraction as a function of time, not rescaled, for the respective correlators.}
\label{fig:ConnCorrs}
\end{figure*}
\begin{figure*}[t]
\includegraphics[width=0.47\linewidth]{\pToFigs/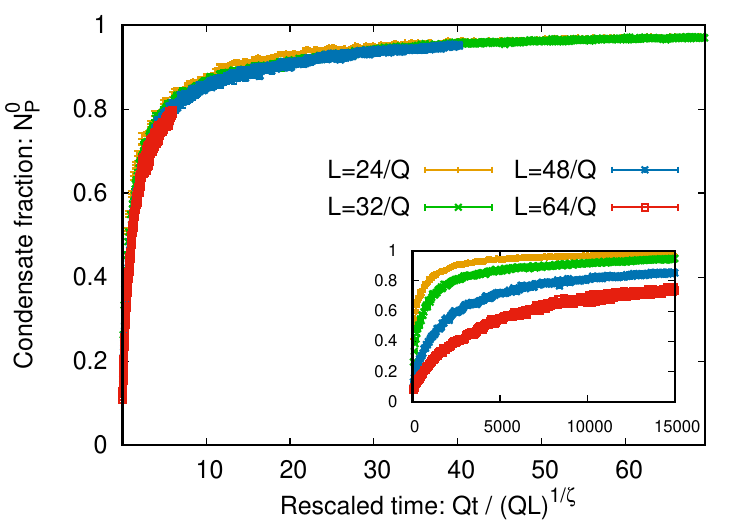}
\includegraphics[width=0.47\linewidth]{\pToFigs/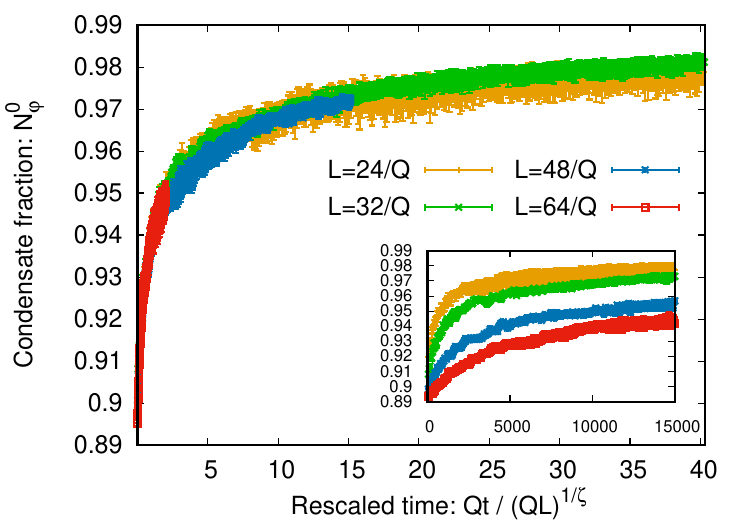}
\caption{Condensate fraction (Eq. \ref{eq:full-corrs}) of the full Polyakov loop correlator (left) and the full algebra field correlator (right) as a function of finite-size rescaled time for different volumes. All curves fall on top of each other and therefore show the emergence of a volume-independent condensate fraction. The scaling exponents $\zeta$ are identical to those in \Cref{fig:ConnCorrs}. Insets: Condensate fraction as a function of time, not rescaled, for the respective correlators. }
\label{fig:full_corrs_scaling}
\end{figure*}
\begin{figure*}[t]
\includegraphics[width=0.47\linewidth]{\pToFigs/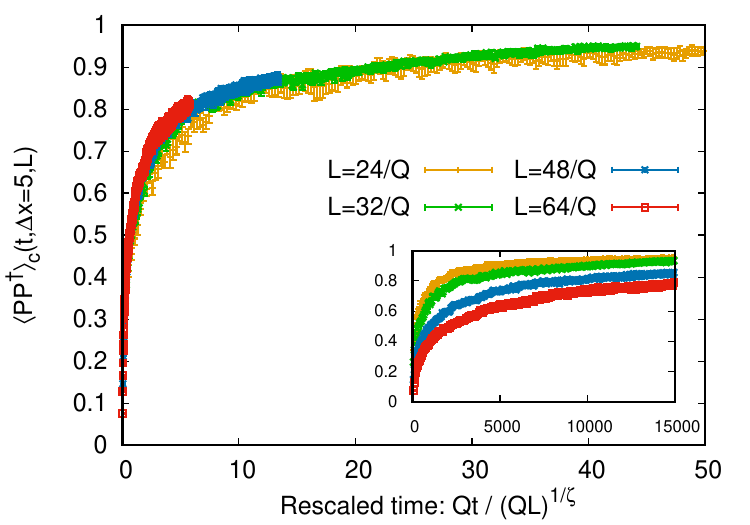}
\includegraphics[width=0.47\linewidth]{\pToFigs/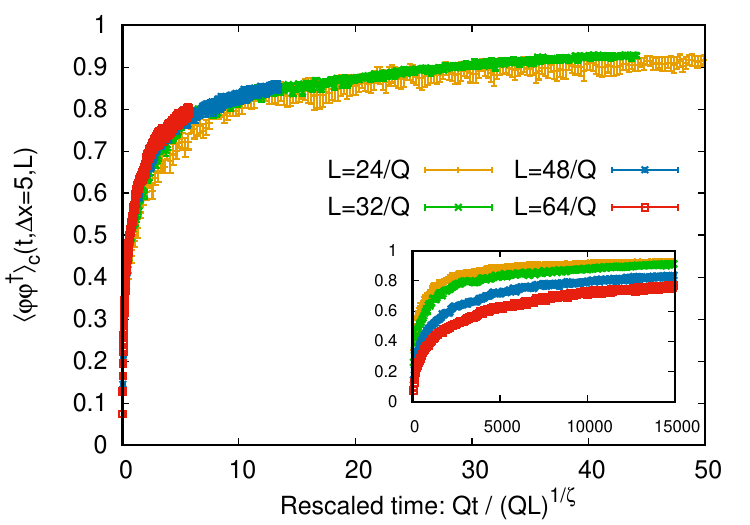}
\caption{Rescaled connected correlators $\langle PP^{\dag}\rangle_c$ and $\CPC$ normalized by their values at $\Delta x = 0$ for fixed $Q \Delta x = 5$, with a reduced time extent ($Qt \leq 15000$ ). The scaling exponents $\zeta$ are identical to those in \Cref{fig:ConnCorrs}. By analyzing the rescaled correlators in this manner, we gain further insights into their temporal evolution and the underlying dynamics of the system under investigation. 
}
\label{fig:spatialcorrscaling}
\end{figure*}
\begin{figure*}[t]
\includegraphics[width=0.49\linewidth]{\pToFigs/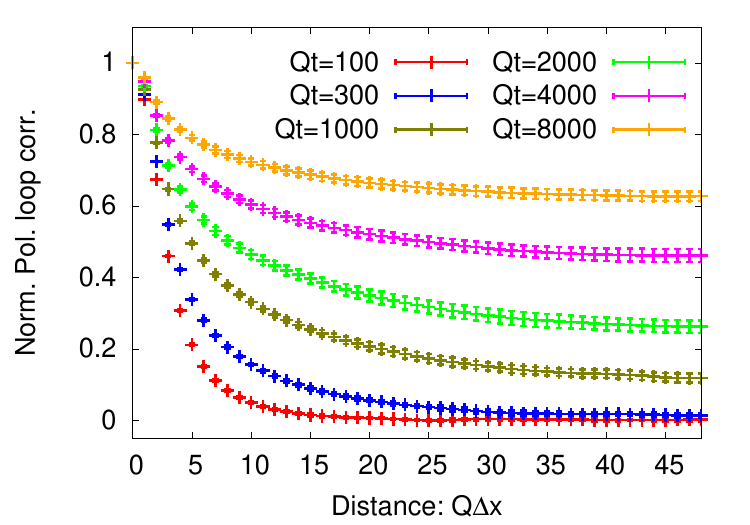}
\includegraphics[width=0.49\linewidth]{\pToFigs/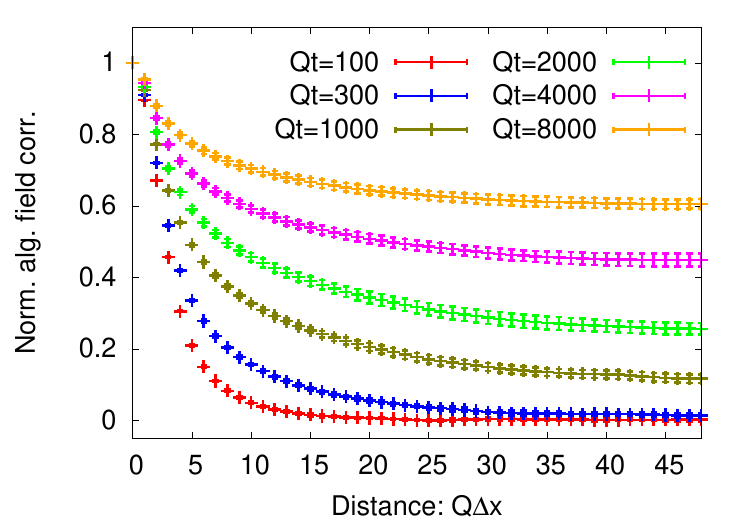}
\caption{Time evolution of the normalized connected Polyakov loop correlator $\langle PP^{\dag}\rangle_c(\Delta x)/\langle PP^{\dag}\rangle_c({\Delta x{=}0})$ (left) and normalized connected algebra field correlator $\langle \varphi \varphi^{\dag}\rangle_c(\Delta x)/\langle \varphi \varphi^{\dag}\rangle_c({\Delta x{=}0})$ (right) shown for six times $Qt=100,300,1000,2000,4000,8000$ on a $N_s=96$ lattice. Clear growth of condensate is demonstrated for both order parameters.
}
\label{fig:time_evo_points}
\end{figure*}
\begin{figure}
\includegraphics[width=\columnwidth]{\pToFigs/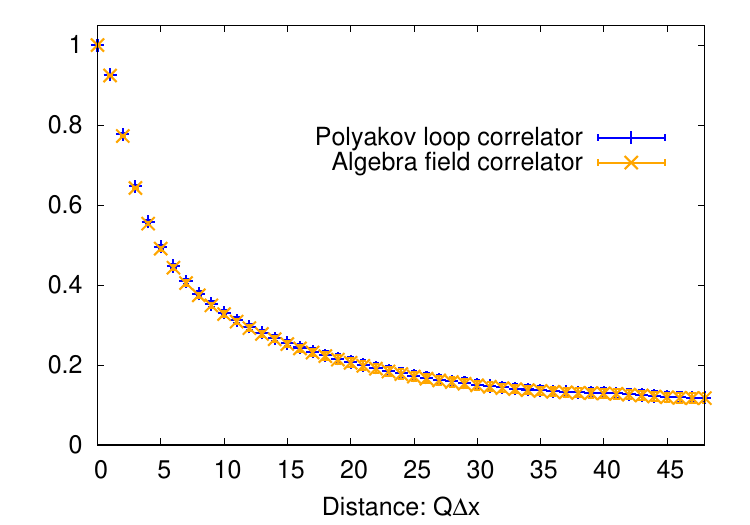}
\caption{Comparative analysis of the normalized connected Polyakov loop correlator $\langle PP^{\dag}\rangle(\Delta x)/\langle PP^{\dag}\rangle({\Delta x{=}0})$ and the normalized connected algebra field correlator $\langle \varphi \varphi^{\dag}\rangle(\Delta x)/\langle \varphi \varphi^{\dag}\rangle({\Delta x{=}0})$ on a lattice with size $N_s=96$ at time $Qt=1000$, as in \Cref{fig:Pol_Wil_illustr} and \Cref{fig:time_evo_points}. The results provide insights into the relationship between these two observables and their behavior in the lattice system under study.
}
\label{fig:PPdag_vs_PhiStarPhi}
\end{figure}
In order to study condensation for the suggested local order parameters, we first consider a general quantity $\mathcal{O}$, for which we can define a condensate fraction. We propose to study the connected two-point correlators, 
\begin{align}
\langle {\cal O}{\cal O}^\dagger \rangle_c (t,\Delta x,L) = \langle {\cal O}{\cal O}^\dagger \rangle - \langle{\cal O}\rangle \langle{\cal O}\rangle^*,
\label{eq:conn-correlators}
\end{align}
where $\mathcal{O}$ is either $P$ or $\varphi$, in order to explicitly distinguish their dynamics from the one-point functions $\langle \mathcal O \rangle$ discussed in the previous subsection.

In one dimension, we define the condensate fraction for a given length $L$ by integrating with respect to the spatial extent $\Delta x$ and dividing out the length $\gamma L$, %
\footnote{We have checked that other powers like $\frac{1}{(\gamma L)^2} \int_0^{\gamma L} d\Delta x\,\Delta x$ lead to similar scaling exponents.}
\begin{align}
    N_{{\cal O}, c}^0 (t,L)&= \frac{1}{\gamma L} \int_0^{\gamma L} d\Delta x\, \frac{\langle {\cal O}{\cal O}^\dagger \rangle_c (t,\Delta x,L)}{\langle {\cal O}{\cal O}^\dagger \rangle_c (t, \Delta x{=}0,L)} ,
    \label{eq:conn-CFs}
\end{align}
where $\gamma < 1$ is a real parameter. The length $L$ in the $x_2$-direction spans the length of the lattice. Since the lattice is periodic, the correlators are symmetric around $\Delta y = L/2$, which constitutes the longest physical distance. In our numerics, we, therefore, use $\gamma=1/4$ and show that our results are consistent with $\gamma=1/2$.

The possibility of interpreting such quantities in \labelcref{eq:conn-CFs} as condensate fractions stems from several observations. Firstly, the integral $\int d \Delta x \langle {\cal O}{\cal O}^\dagger \rangle_c (t,\Delta x, L)$ resembles a Fourier transformation, or rather Wigner transform for a vanishing momentum $p=0$. 
Since $\langle {\cal O}{\cal O}^\dagger \rangle_c (t, \Delta x{=}0, L)$ can be interpreted as an integral over all momentum modes $\int dp \langle {\cal O}{\cal O}^\dagger \rangle_c (t, p, L)$ by Fourier transforming $\langle {\cal O}{\cal O}^\dagger \rangle_c (t,\Delta x, L)$, dividing the zero mode by this quantity corresponds to its fraction. 
Finally, dividing by a suitable volume or length element bounds it to unity and allows one to interpret this zero-mode fraction as a condensate if it becomes large and independent of the lattice size. However, such lattice size independence is difficult to show since the condensate fraction keeps growing due to the far-from-equilibrium dynamics. Therefore we define a volume-dependent condensate formation time in a similar way as for Wilson loops in \cite{Berges:2019oun} and relativistic and nonrelativistic scalar field theories in \cite{Orioli:2015dxa}, as
\begin{align}
 \label{eq:cond_time}
 \tcond = (QL)^{1/\zeta}.
\end{align}
Rescaling time by $t / \tcond$ gets rid of the remaining volume dependence in the evolution of the condensate fractions $N_{{\cal O}, c}^0$, as we will demonstrate numerically below, finally signaling the emergence of a condensate.
In contrast, an interpretation in terms of a condensate is not possible if this quantity remains volume dependent. This is for instance the case in thermal equilibrium. We note that this particular form of condensate formation is a genuinely far-from-equilibrium phenomenon, usually requires large occupation numbers, and is often associated with an inverse particle cascade that occupies the zero mode, as observed for scalar theories \cite{Orioli:2015dxa, Schmied:2018mte}.

In \Cref{fig:ConnCorrs}, we demonstrate the emergence of a volume-independent condensate fraction for both the connected Polyakov loop (left) and the connected algebra field correlators (right) computed using \labelcref{eq:conn-CFs}. 
In particular, we show $N_{P, c}^0$ and $N_{\varphi, c}^0$ for four different volumes as functions of time $Q t$ in the insets, which yields a volume-dependent evolution. In contrast, rescaling time by $\tcond$, leads to volume-independent curves, as visible in the main panels. 
The initial growth of the zero mode necessarily ceases as the volume becomes correlated; this is demonstrated in the figure as the zero mode grows with time for each lattice size and then levels off at the time scale $\sim \tcond$. 

The time scaling is characterized by the scaling exponent $\zeta$ entering $\tcond$ in \labelcref{eq:cond_time}. We find for the condensate fraction of the connected Polyakov loop correlator, $\zeta=0.31$, and for the connected algebra field correlator, $\zeta=0.34$. The collapse of the curves onto one another in \Cref{fig:ConnCorrs} illustrates the volume-independent nature of the condensation phenomenon. 

Although the dynamics is captured mainly by the connected correlators, we also check the condensate fraction scaling in terms of the full two-point functions of the spatial Polyakov loop and the algebra field in \Cref{fig:full_corrs_scaling}. For a generic order parameter $\mathcal{O}$, this condensate fraction is given as:
\begin{align}
    N_{\cal O}^0(t,L)&= \frac{1}{\gamma L} \int_0^{\gamma L} d\Delta x\, \frac{\langle {\cal O}{\cal O}^\dagger \rangle(t,\Delta x,L)}{\langle {\cal O}{\cal O}^\dagger \rangle (t, \Delta x{=}0,L)}.
\label{eq:full-corrs}
\end{align}
The condensate fractions for the full correlators scale with similar exponents $\zeta$ as their connected parts. 

In \Cref{tab:exponents-summary}, we summarize the scaling exponents for the connected correlators and compare them to the scaling results of $N_{P}^0(t,L)$, and to those found in our previous publication \cite{Berges:2019oun} for the Wilson loop expectation value using our conventions. We note that the latter two quantities are intrinsically connected to each other but are not the same, as shown in \Cref{fig:DepictionPol_Wil} and \Cref{fig:Pol_Wil_illustr}. We observe that all of the extracted scaling exponents agree within statistical error, while showing a slight systematic dependence on the maximal distance $\Delta x$ in the integrals. For $L/4$ the value for $\zeta$ tends to be around $10 - 15\%$ smaller than for $L/2$. 

We then check the scaling of the connected correlators $\langle P P^\dag \rangle$ and $\langle \varphi \varphi^\dag \rangle$ in \labelcref{eq:conn-correlators} at a fixed $\Delta x$, as shown in \Cref{fig:spatialcorrscaling}. Interestingly, both the Polyakov loop and the algebra field correlators show the same scaling as their integrated counterparts in \Cref{fig:ConnCorrs}. We find the values $\zeta=0.31$ for the connected Polyakov loop correlator and $\zeta=0.34$ for the connected algebra field correlator. This implies that not only correlations at large distances grow in this self-similar way, but also at finite distances in a similar fashion. 

To further visualize this condensation phenomenon, we show the time evolution for the connected Polyakov loop and algebra field correlators for a lattice of size $N_s=96$ in \Cref{fig:time_evo_points}. The connected correlators are plotted as functions of the distance $\Delta x$ for six different times. One observes that for early times $\Q t = 100$ and $300$, both correlations are zero are large distances, signaling no condensate. However, at $Q t \gtrsim 1000$ a plateau has started to form at all distances, which further grows over time. This demonstrates a phase transition between a phase without and one with a condensate, where the condensate formation time scales with volume according to $\tcond$ in \labelcref{eq:cond_time}. 

Interestingly, these connected correlators $\langle PP^{\dag}\rangle(\Delta x)/\langle PP^{\dag}\rangle({\Delta x{=}0})$ and $\langle \varphi \varphi^{\dag}\rangle(\Delta x)/\langle \varphi \varphi^{\dag}\rangle({\Delta x{=}0})$ agree even quantitatively, as shown in 
\Cref{fig:PPdag_vs_PhiStarPhi}. A similar agreement is known from (non-relativistic) scalar models \cite{Mikheev:2018adp}. There field correlations associated to the particle number can be dominated by excitations of the fluctuating phase-angle fields at sufficiently low momenta during Bose-Einstein condensation far from equilibrium. This observation provides another interesting analogy to scalar systems.

\section{Conclusion}
\label{sec:conclusion}

In this study we have demonstrated the existence of gauge condensation in over-occupied QCD plasmas. Two order parameters related to the Wilson loop demonstrate the build up of a macroscopic zero mode characteristic of condensation phenomena. These order parameters are the spatial Polyakov loop and algebra-valued scalar holonomous eigenvalue field, which form connected two-point correlation functions of which the condensate is obtained from their zero modes. 

Through classical-statistical lattice simulations of the Yang-Mills plasma with large gluon densities and weak coupling, we have observed the emergence of condensate fractions on time scales $\tcond \sim L^{1/\zeta}$. The extracted values for the universal scaling exponent $\zeta$ are consistent with the different correlators employed and also agree with previous studies for the spatial Wilson loop within error. We have shown that the growth of the corresponding zero modes can also be seen in the correlations at large distances and is independent of the dynamics of the one-point functions of the order parameters. 

The algebra-valued local scalar field can be related to scalar Bose condensation. Furthermore, the use of a local scalar field allows us to construct effective actions in terms of the algebra field, which is naturally suited to describe infrared dynamics far from equilibrium. We have constructed a suitable effective action and leave the connection to kinetic or other approaches to future studies.

The findings presented in this study contribute to the ongoing quest for a deeper understanding of the non-equilibrium dynamics of QCD matter. Moreover, they provide valuable insights into the challenges associated with defining gauge condensation and offer new perspectives on the early-time evolution of heavy-ion collisions. This work further paves the way for investigations into the rich physics of non-equilibrium QCD and its connections to other areas of theoretical and experimental physics.

\begin{acknowledgements}

We thank M.~Mace, R.~Pisarski, P.~Radpay, and R.~Venugopalan for discussions. 
This work is funded by the Deutsche Forschungsgemeinschaft (DFG, German Research Foundation) by the Collaborative Research Centre SFB 1225 - 273811115 (ISOQUANT) and Germany’s Excellence Strategy EXC 2181/1 - 390900948 (the Heidelberg STRUCTURES Excellence Cluster). KB would also like to thank the Austrian Science Fund (FWF) for support under project P 34455. 
The computational results presented have been achieved using the Vienna Scientific Cluster (VSC), project 71444, and bwUniCluster (2.0), supported by the state of Baden-Württemberg through bwHPC.

\end{acknowledgements}

\appendix

\section{Mean and error estimates for $\zeta$}
\label{app:zeta_estimate}

Let us consider a time and lattice-size dependent quantity $A(t,L)$ like for instance $\langle {\cal O} \rangle(t, L)$ or $N_{{\cal O}, c}^0 (t,L)$. We assume that its dependence can be reduced to a single argument that combines $t$ and $L$ using $\tcond = (QL)^{1/\zeta}$ as in \labelcref{eq:cond_time}. Our goal is to estimate the most likely value for the exponent $\zeta$ and its uncertainty such that the quantity $A$ becomes a function of $t/\tcond$ only. For this, we use an adapted version of the $\chi^2$-procedure in \cite{Berges:2013fga, Orioli:2015dxa, Berges:2019oun}. 

In particular, we choose the lattice lengths $L_i\in\{24, 32, 48, 64\}$, with $L = N_s a_s$. We drop the transient early time evolution before scaling sets in considering only $t > \{ 100, 300, 300, 600 \}$, respectively, interpolate our data and smoothen it using a Savitzky-Golay filter. Next, we vary $\zeta$ within $[0.2, 0.6]$ with $0.005$ steps, rescale time using $\tcond$, which leads to $A_i(t, \zeta) \equiv A(t/\tcond(L_i, \zeta), L_i)$, and define 
\begin{align}
    \label{eq:chisqr}
    \chi^2(\zeta)=\frac{1}{5} \sum_{(i,j)} \frac{\int dt (A_i(t,\zeta)-A_j(t,\zeta))^2}{\int dt (A_i(t,\zeta))^2}.
\end{align}
Here, the tuple $(i,j)$ runs over five pairs of lattice sizes which are explicitly chosen as $(i,j)\in\{(1,2),(1,3),(2,3),(2,4),(3,4)\}$, and $\int dt$ integrates over a shared domain of the rescaled times of $A_i$. 

We approximate the likelihood distribution of $\zeta$ as
\begin{align}
    P(\zeta)= \exp \left\{ -\frac{\chi^2(\zeta)}{2\chi^2_{min}} \right\},
\end{align}
with $\chi^2_{\mathrm{min}}$ being the minimum value of $\chi^2$ in \labelcref{eq:chisqr}. 
Finally, the optimal value of the exponent $\bar\zeta$ and its uncertainty $\sigma_\zeta$ are estimated by fitting a Gaussian function of the form
\begin{align}
    f(\bar\zeta,\sigma_\zeta,\zeta)=\mathcal{N}\exp\left\{-\frac{(\zeta_{opt}-\zeta)^2}{2 \sigma_\zeta^2}\right\}
\end{align}
to the likelihood distribution $P(\zeta)$. Our results in \Cref{tab:exponents-summary} are given by $\zeta = \bar\zeta \pm \sigma_\zeta$.

\bibliography{Bibliography}

\begin{thebibliography}{40}%
\makeatletter
\providecommand \@ifxundefined [1]{%
 \@ifx{#1\undefined}
}%
\providecommand \@ifnum [1]{%
 \ifnum #1\expandafter \@firstoftwo
 \else \expandafter \@secondoftwo
 \fi
}%
\providecommand \@ifx [1]{%
 \ifx #1\expandafter \@firstoftwo
 \else \expandafter \@secondoftwo
 \fi
}%
\providecommand \natexlab [1]{#1}%
\providecommand \enquote  [1]{``#1''}%
\providecommand \bibnamefont  [1]{#1}%
\providecommand \bibfnamefont [1]{#1}%
\providecommand \citenamefont [1]{#1}%
\providecommand \href@noop [0]{\@secondoftwo}%
\providecommand \href [0]{\begingroup \@sanitize@url \@href}%
\providecommand \@href[1]{\@@startlink{#1}\@@href}%
\providecommand \@@href[1]{\endgroup#1\@@endlink}%
\providecommand \@sanitize@url [0]{\catcode `\\12\catcode `\$12\catcode
  `\&12\catcode `\#12\catcode `\^12\catcode `\_12\catcode `\%12\relax}%
\providecommand \@@startlink[1]{}%
\providecommand \@@endlink[0]{}%
\providecommand \url  [0]{\begingroup\@sanitize@url \@url }%
\providecommand \@url [1]{\endgroup\@href {#1}{\urlprefix }}%
\providecommand \urlprefix  [0]{URL }%
\providecommand \Eprint [0]{\href }%
\providecommand \doibase [0]{http://dx.doi.org/}%
\providecommand \selectlanguage [0]{\@gobble}%
\providecommand \bibinfo  [0]{\@secondoftwo}%
\providecommand \bibfield  [0]{\@secondoftwo}%
\providecommand \translation [1]{[#1]}%
\providecommand \BibitemOpen [0]{}%
\providecommand \bibitemStop [0]{}%
\providecommand \bibitemNoStop [0]{.\EOS\space}%
\providecommand \EOS [0]{\spacefactor3000\relax}%
\providecommand \BibitemShut  [1]{\csname bibitem#1\endcsname}%
\let\auto@bib@innerbib\@empty
\bibitem [{\citenamefont {Gelis}\ \emph {et~al.}(2010)\citenamefont {Gelis},
  \citenamefont {Iancu}, \citenamefont {Jalilian-Marian},\ and\ \citenamefont
  {Venugopalan}}]{Gelis:2010nm}%
  \BibitemOpen
  \bibfield  {author} {\bibinfo {author} {\bibfnamefont {F.}~\bibnamefont
  {Gelis}}, \bibinfo {author} {\bibfnamefont {E.}~\bibnamefont {Iancu}},
  \bibinfo {author} {\bibfnamefont {J.}~\bibnamefont {Jalilian-Marian}}, \ and\
  \bibinfo {author} {\bibfnamefont {R.}~\bibnamefont {Venugopalan}},\ }\href
  {\doibase 10.1146/annurev.nucl.010909.083629} {\bibfield  {journal} {\bibinfo
   {journal} {Ann. Rev. Nucl. Part. Sci.}\ }\textbf {\bibinfo {volume} {60}},\
  \bibinfo {pages} {463} (\bibinfo {year} {2010})}\BibitemShut {NoStop}%
\bibitem [{\citenamefont {Lappi}\ and\ \citenamefont
  {McLerran}(2006)}]{Lappi:2006fp}%
  \BibitemOpen
  \bibfield  {author} {\bibinfo {author} {\bibfnamefont {T.}~\bibnamefont
  {Lappi}}\ and\ \bibinfo {author} {\bibfnamefont {L.}~\bibnamefont
  {McLerran}},\ }\href {\doibase 10.1016/j.nuclphysa.2006.04.001} {\bibfield
  {journal} {\bibinfo  {journal} {Nucl. Phys.}\ }\textbf {\bibinfo {volume}
  {A772}},\ \bibinfo {pages} {200} (\bibinfo {year} {2006})}\BibitemShut
  {NoStop}%
\bibitem [{\citenamefont {Blaizot}\ \emph {et~al.}(2012)\citenamefont
  {Blaizot}, \citenamefont {Gelis}, \citenamefont {Liao}, \citenamefont
  {McLerran},\ and\ \citenamefont {Venugopalan}}]{Blaizot:2011xf}%
  \BibitemOpen
  \bibfield  {author} {\bibinfo {author} {\bibfnamefont {J.-P.}\ \bibnamefont
  {Blaizot}}, \bibinfo {author} {\bibfnamefont {F.}~\bibnamefont {Gelis}},
  \bibinfo {author} {\bibfnamefont {J.-F.}\ \bibnamefont {Liao}}, \bibinfo
  {author} {\bibfnamefont {L.}~\bibnamefont {McLerran}}, \ and\ \bibinfo
  {author} {\bibfnamefont {R.}~\bibnamefont {Venugopalan}},\ }\href {\doibase
  10.1016/j.nuclphysa.2011.10.005} {\bibfield  {journal} {\bibinfo  {journal}
  {Nucl. Phys.}\ }\textbf {\bibinfo {volume} {A873}},\ \bibinfo {pages} {68}
  (\bibinfo {year} {2012})}\BibitemShut {NoStop}%
\bibitem [{\citenamefont {Kurkela}\ and\ \citenamefont
  {Moore}(2012)}]{Kurkela:2012hp}%
  \BibitemOpen
  \bibfield  {author} {\bibinfo {author} {\bibfnamefont {A.}~\bibnamefont
  {Kurkela}}\ and\ \bibinfo {author} {\bibfnamefont {G.~D.}\ \bibnamefont
  {Moore}},\ }\href {\doibase 10.1103/PhysRevD.86.056008} {\bibfield  {journal}
  {\bibinfo  {journal} {Phys. Rev.}\ }\textbf {\bibinfo {volume} {D86}},\
  \bibinfo {pages} {056008} (\bibinfo {year} {2012})}\BibitemShut {NoStop}%
\bibitem [{\citenamefont {Berges}\ \emph
  {et~al.}(2014{\natexlab{a}})\citenamefont {Berges}, \citenamefont
  {Boguslavski}, \citenamefont {Schlichting},\ and\ \citenamefont
  {Venugopalan}}]{Berges:2013eia}%
  \BibitemOpen
  \bibfield  {author} {\bibinfo {author} {\bibfnamefont {J.}~\bibnamefont
  {Berges}}, \bibinfo {author} {\bibfnamefont {K.}~\bibnamefont {Boguslavski}},
  \bibinfo {author} {\bibfnamefont {S.}~\bibnamefont {Schlichting}}, \ and\
  \bibinfo {author} {\bibfnamefont {R.}~\bibnamefont {Venugopalan}},\ }\href
  {\doibase 10.1103/PhysRevD.89.074011} {\bibfield  {journal} {\bibinfo
  {journal} {Phys. Rev.}\ }\textbf {\bibinfo {volume} {D89}},\ \bibinfo {pages}
  {074011} (\bibinfo {year} {2014}{\natexlab{a}})}\BibitemShut {NoStop}%
\bibitem [{\citenamefont {Abraao~York}\ \emph {et~al.}(2014)\citenamefont
  {Abraao~York}, \citenamefont {Kurkela}, \citenamefont {Lu},\ and\
  \citenamefont {Moore}}]{AbraaoYork:2014hbk}%
  \BibitemOpen
  \bibfield  {author} {\bibinfo {author} {\bibfnamefont {M.~C.}\ \bibnamefont
  {Abraao~York}}, \bibinfo {author} {\bibfnamefont {A.}~\bibnamefont
  {Kurkela}}, \bibinfo {author} {\bibfnamefont {E.}~\bibnamefont {Lu}}, \ and\
  \bibinfo {author} {\bibfnamefont {G.~D.}\ \bibnamefont {Moore}},\ }\href
  {\doibase 10.1103/PhysRevD.89.074036} {\bibfield  {journal} {\bibinfo
  {journal} {Phys. Rev. D}\ }\textbf {\bibinfo {volume} {89}},\ \bibinfo
  {pages} {074036} (\bibinfo {year} {2014})},\ \Eprint
  {http://arxiv.org/abs/1401.3751} {arXiv:1401.3751 [hep-ph]} \BibitemShut
  {NoStop}%
\bibitem [{\citenamefont {Blaizot}\ \emph {et~al.}(2017)\citenamefont
  {Blaizot}, \citenamefont {Liao},\ and\ \citenamefont
  {Mehtar-Tani}}]{Blaizot:2016iir}%
  \BibitemOpen
  \bibfield  {author} {\bibinfo {author} {\bibfnamefont {J.-P.}\ \bibnamefont
  {Blaizot}}, \bibinfo {author} {\bibfnamefont {J.}~\bibnamefont {Liao}}, \
  and\ \bibinfo {author} {\bibfnamefont {Y.}~\bibnamefont {Mehtar-Tani}},\
  }\href {\doibase 10.1016/j.nuclphysa.2017.02.003} {\bibfield  {journal}
  {\bibinfo  {journal} {Nucl. Phys.}\ }\textbf {\bibinfo {volume} {A961}},\
  \bibinfo {pages} {37} (\bibinfo {year} {2017})}\BibitemShut {NoStop}%
\bibitem [{\citenamefont {Gasenzer}\ \emph {et~al.}(2014)\citenamefont
  {Gasenzer}, \citenamefont {McLerran}, \citenamefont {Pawlowski},\ and\
  \citenamefont {Sexty}}]{Gasenzer:2013era}%
  \BibitemOpen
  \bibfield  {author} {\bibinfo {author} {\bibfnamefont {T.}~\bibnamefont
  {Gasenzer}}, \bibinfo {author} {\bibfnamefont {L.}~\bibnamefont {McLerran}},
  \bibinfo {author} {\bibfnamefont {J.~M.}\ \bibnamefont {Pawlowski}}, \ and\
  \bibinfo {author} {\bibfnamefont {D.}~\bibnamefont {Sexty}},\ }\href
  {\doibase 10.1016/j.nuclphysa.2014.07.030} {\bibfield  {journal} {\bibinfo
  {journal} {Nucl. Phys.}\ }\textbf {\bibinfo {volume} {A930}},\ \bibinfo
  {pages} {163} (\bibinfo {year} {2014})}\BibitemShut {NoStop}%
\bibitem [{\citenamefont {Ford}\ \emph {et~al.}(1998)\citenamefont {Ford},
  \citenamefont {Mitreuter}, \citenamefont {Tok}, \citenamefont {Wipf},\ and\
  \citenamefont {Pawlowski}}]{Ford:1998bt}%
  \BibitemOpen
  \bibfield  {author} {\bibinfo {author} {\bibfnamefont {C.}~\bibnamefont
  {Ford}}, \bibinfo {author} {\bibfnamefont {U.~G.}\ \bibnamefont {Mitreuter}},
  \bibinfo {author} {\bibfnamefont {T.}~\bibnamefont {Tok}}, \bibinfo {author}
  {\bibfnamefont {A.}~\bibnamefont {Wipf}}, \ and\ \bibinfo {author}
  {\bibfnamefont {J.~M.}\ \bibnamefont {Pawlowski}},\ }\href {\doibase
  10.1006/aphy.1998.5841} {\bibfield  {journal} {\bibinfo  {journal} {Annals
  Phys.}\ }\textbf {\bibinfo {volume} {269}},\ \bibinfo {pages} {26} (\bibinfo
  {year} {1998})}\BibitemShut {NoStop}%
\bibitem [{\citenamefont {Mitreuter}\ \emph {et~al.}(1998)\citenamefont
  {Mitreuter}, \citenamefont {Pawlowski},\ and\ \citenamefont
  {Wipf}}]{Mitreuter:1996ze}%
  \BibitemOpen
  \bibfield  {author} {\bibinfo {author} {\bibfnamefont {U.~G.}\ \bibnamefont
  {Mitreuter}}, \bibinfo {author} {\bibfnamefont {J.~M.}\ \bibnamefont
  {Pawlowski}}, \ and\ \bibinfo {author} {\bibfnamefont {A.}~\bibnamefont
  {Wipf}},\ }\href {\doibase 10.1016/S0550-3213(97)00733-5} {\bibfield
  {journal} {\bibinfo  {journal} {Nucl. Phys.}\ }\textbf {\bibinfo {volume}
  {B514}},\ \bibinfo {pages} {381} (\bibinfo {year} {1998})}\BibitemShut
  {NoStop}%
\bibitem [{\citenamefont {Berges}\ \emph {et~al.}(2020)\citenamefont {Berges},
  \citenamefont {Boguslavski}, \citenamefont {Mace},\ and\ \citenamefont
  {Pawlowski}}]{Berges:2019oun}%
  \BibitemOpen
  \bibfield  {author} {\bibinfo {author} {\bibfnamefont {J.}~\bibnamefont
  {Berges}}, \bibinfo {author} {\bibfnamefont {K.}~\bibnamefont {Boguslavski}},
  \bibinfo {author} {\bibfnamefont {M.}~\bibnamefont {Mace}}, \ and\ \bibinfo
  {author} {\bibfnamefont {J.~M.}\ \bibnamefont {Pawlowski}},\ }\href {\doibase
  10.1103/PhysRevD.102.034014} {\bibfield  {journal} {\bibinfo  {journal}
  {Phys. Rev. D}\ }\textbf {\bibinfo {volume} {102}},\ \bibinfo {pages}
  {034014} (\bibinfo {year} {2020})},\ \Eprint
  {http://arxiv.org/abs/1909.06147} {arXiv:1909.06147 [hep-ph]} \BibitemShut
  {NoStop}%
\bibitem [{\citenamefont {Berges}\ and\ \citenamefont
  {Sexty}(2012)}]{Berges:2012us}%
  \BibitemOpen
  \bibfield  {author} {\bibinfo {author} {\bibfnamefont {J.}~\bibnamefont
  {Berges}}\ and\ \bibinfo {author} {\bibfnamefont {D.}~\bibnamefont {Sexty}},\
  }\href {\doibase 10.1103/PhysRevLett.108.161601} {\bibfield  {journal}
  {\bibinfo  {journal} {Phys. Rev. Lett.}\ }\textbf {\bibinfo {volume} {108}},\
  \bibinfo {pages} {161601} (\bibinfo {year} {2012})}\BibitemShut {NoStop}%
\bibitem [{\citenamefont {Berges}\ \emph {et~al.}(2008)\citenamefont {Berges},
  \citenamefont {Scheffler},\ and\ \citenamefont {Sexty}}]{Berges:2007re}%
  \BibitemOpen
  \bibfield  {author} {\bibinfo {author} {\bibfnamefont {J.}~\bibnamefont
  {Berges}}, \bibinfo {author} {\bibfnamefont {S.}~\bibnamefont {Scheffler}}, \
  and\ \bibinfo {author} {\bibfnamefont {D.}~\bibnamefont {Sexty}},\ }\href
  {\doibase 10.1103/PhysRevD.77.034504} {\bibfield  {journal} {\bibinfo
  {journal} {Phys. Rev.}\ }\textbf {\bibinfo {volume} {D77}},\ \bibinfo {pages}
  {034504} (\bibinfo {year} {2008})}\BibitemShut {NoStop}%
\bibitem [{\citenamefont {Dumitru}\ \emph {et~al.}(2014)\citenamefont
  {Dumitru}, \citenamefont {Lappi},\ and\ \citenamefont
  {Nara}}]{Dumitru:2014nka}%
  \BibitemOpen
  \bibfield  {author} {\bibinfo {author} {\bibfnamefont {A.}~\bibnamefont
  {Dumitru}}, \bibinfo {author} {\bibfnamefont {T.}~\bibnamefont {Lappi}}, \
  and\ \bibinfo {author} {\bibfnamefont {Y.}~\bibnamefont {Nara}},\ }\href
  {\doibase 10.1016/j.physletb.2014.05.005} {\bibfield  {journal} {\bibinfo
  {journal} {Phys. Lett.}\ }\textbf {\bibinfo {volume} {B734}},\ \bibinfo
  {pages} {7} (\bibinfo {year} {2014})}\BibitemShut {NoStop}%
\bibitem [{\citenamefont {Mace}\ \emph {et~al.}(2016)\citenamefont {Mace},
  \citenamefont {Schlichting},\ and\ \citenamefont
  {Venugopalan}}]{Mace:2016svc}%
  \BibitemOpen
  \bibfield  {author} {\bibinfo {author} {\bibfnamefont {M.}~\bibnamefont
  {Mace}}, \bibinfo {author} {\bibfnamefont {S.}~\bibnamefont {Schlichting}}, \
  and\ \bibinfo {author} {\bibfnamefont {R.}~\bibnamefont {Venugopalan}},\
  }\href {\doibase 10.1103/PhysRevD.93.074036} {\bibfield  {journal} {\bibinfo
  {journal} {Phys. Rev.}\ }\textbf {\bibinfo {volume} {D93}},\ \bibinfo {pages}
  {074036} (\bibinfo {year} {2016})}\BibitemShut {NoStop}%
\bibitem [{\citenamefont {Berges}\ \emph {et~al.}(2017)\citenamefont {Berges},
  \citenamefont {Mace},\ and\ \citenamefont {Schlichting}}]{Berges:2017igc}%
  \BibitemOpen
  \bibfield  {author} {\bibinfo {author} {\bibfnamefont {J.}~\bibnamefont
  {Berges}}, \bibinfo {author} {\bibfnamefont {M.}~\bibnamefont {Mace}}, \ and\
  \bibinfo {author} {\bibfnamefont {S.}~\bibnamefont {Schlichting}},\ }\href
  {\doibase 10.1103/PhysRevLett.118.192005} {\bibfield  {journal} {\bibinfo
  {journal} {Phys. Rev. Lett.}\ }\textbf {\bibinfo {volume} {118}},\ \bibinfo
  {pages} {192005} (\bibinfo {year} {2017})}\BibitemShut {NoStop}%
\bibitem [{\citenamefont {Berges}\ \emph
  {et~al.}(2014{\natexlab{b}})\citenamefont {Berges}, \citenamefont
  {Boguslavski}, \citenamefont {Schlichting},\ and\ \citenamefont
  {Venugopalan}}]{Berges:2013fga}%
  \BibitemOpen
  \bibfield  {author} {\bibinfo {author} {\bibfnamefont {J.}~\bibnamefont
  {Berges}}, \bibinfo {author} {\bibfnamefont {K.}~\bibnamefont {Boguslavski}},
  \bibinfo {author} {\bibfnamefont {S.}~\bibnamefont {Schlichting}}, \ and\
  \bibinfo {author} {\bibfnamefont {R.}~\bibnamefont {Venugopalan}},\ }\href
  {\doibase 10.1103/PhysRevD.89.114007} {\bibfield  {journal} {\bibinfo
  {journal} {Phys. Rev.}\ }\textbf {\bibinfo {volume} {D89}},\ \bibinfo {pages}
  {114007} (\bibinfo {year} {2014}{\natexlab{b}})}\BibitemShut {NoStop}%
\bibitem [{\citenamefont {Berges}\ \emph
  {et~al.}(2015{\natexlab{a}})\citenamefont {Berges}, \citenamefont
  {Boguslavski}, \citenamefont {Schlichting},\ and\ \citenamefont
  {Venugopalan}}]{Berges:2014bba}%
  \BibitemOpen
  \bibfield  {author} {\bibinfo {author} {\bibfnamefont {J.}~\bibnamefont
  {Berges}}, \bibinfo {author} {\bibfnamefont {K.}~\bibnamefont {Boguslavski}},
  \bibinfo {author} {\bibfnamefont {S.}~\bibnamefont {Schlichting}}, \ and\
  \bibinfo {author} {\bibfnamefont {R.}~\bibnamefont {Venugopalan}},\ }\href
  {\doibase 10.1103/PhysRevLett.114.061601} {\bibfield  {journal} {\bibinfo
  {journal} {Phys. Rev. Lett.}\ }\textbf {\bibinfo {volume} {114}},\ \bibinfo
  {pages} {061601} (\bibinfo {year} {2015}{\natexlab{a}})}\BibitemShut
  {NoStop}%
\bibitem [{\citenamefont {Berges}\ \emph
  {et~al.}(2015{\natexlab{b}})\citenamefont {Berges}, \citenamefont
  {Boguslavski}, \citenamefont {Schlichting},\ and\ \citenamefont
  {Venugopalan}}]{Berges:2015ixa}%
  \BibitemOpen
  \bibfield  {author} {\bibinfo {author} {\bibfnamefont {J.}~\bibnamefont
  {Berges}}, \bibinfo {author} {\bibfnamefont {K.}~\bibnamefont {Boguslavski}},
  \bibinfo {author} {\bibfnamefont {S.}~\bibnamefont {Schlichting}}, \ and\
  \bibinfo {author} {\bibfnamefont {R.}~\bibnamefont {Venugopalan}},\ }\href
  {\doibase 10.1103/PhysRevD.92.096006} {\bibfield  {journal} {\bibinfo
  {journal} {Phys. Rev.}\ }\textbf {\bibinfo {volume} {D92}},\ \bibinfo {pages}
  {096006} (\bibinfo {year} {2015}{\natexlab{b}})}\BibitemShut {NoStop}%
\bibitem [{\citenamefont {Piñeiro~Orioli}\ \emph {et~al.}(2015)\citenamefont
  {Piñeiro~Orioli}, \citenamefont {Boguslavski},\ and\ \citenamefont
  {Berges}}]{Orioli:2015dxa}%
  \BibitemOpen
  \bibfield  {author} {\bibinfo {author} {\bibfnamefont {A.}~\bibnamefont
  {Piñeiro~Orioli}}, \bibinfo {author} {\bibfnamefont {K.}~\bibnamefont
  {Boguslavski}}, \ and\ \bibinfo {author} {\bibfnamefont {J.}~\bibnamefont
  {Berges}},\ }\href {\doibase 10.1103/PhysRevD.92.025041} {\bibfield
  {journal} {\bibinfo  {journal} {Phys. Rev.}\ }\textbf {\bibinfo {volume}
  {D92}},\ \bibinfo {pages} {025041} (\bibinfo {year} {2015})}\BibitemShut
  {NoStop}%
\bibitem [{\citenamefont {Schachner}\ \emph {et~al.}(2017)\citenamefont
  {Schachner}, \citenamefont {Pi\~{n}eiro Orioli},\ and\ \citenamefont
  {Berges}}]{Schachner:2016frd}%
  \BibitemOpen
  \bibfield  {author} {\bibinfo {author} {\bibfnamefont {A.}~\bibnamefont
  {Schachner}}, \bibinfo {author} {\bibfnamefont {A.}~\bibnamefont {Pi\~{n}eiro
  Orioli}}, \ and\ \bibinfo {author} {\bibfnamefont {J.}~\bibnamefont
  {Berges}},\ }\href {\doibase 10.1103/PhysRevA.95.053605} {\bibfield
  {journal} {\bibinfo  {journal} {Phys. Rev.}\ }\textbf {\bibinfo {volume}
  {A95}},\ \bibinfo {pages} {053605} (\bibinfo {year} {2017})}\BibitemShut
  {NoStop}%
\bibitem [{\citenamefont {Walz}\ \emph {et~al.}(2018)\citenamefont {Walz},
  \citenamefont {Boguslavski},\ and\ \citenamefont {Berges}}]{Walz:2017ffj}%
  \BibitemOpen
  \bibfield  {author} {\bibinfo {author} {\bibfnamefont {R.}~\bibnamefont
  {Walz}}, \bibinfo {author} {\bibfnamefont {K.}~\bibnamefont {Boguslavski}}, \
  and\ \bibinfo {author} {\bibfnamefont {J.}~\bibnamefont {Berges}},\ }\href
  {\doibase 10.1103/PhysRevD.97.116011} {\bibfield  {journal} {\bibinfo
  {journal} {Phys. Rev.}\ }\textbf {\bibinfo {volume} {D97}},\ \bibinfo {pages}
  {116011} (\bibinfo {year} {2018})}\BibitemShut {NoStop}%
\bibitem [{\citenamefont {Chantesana}\ \emph {et~al.}(2019)\citenamefont
  {Chantesana}, \citenamefont {Piñeiro~Orioli},\ and\ \citenamefont
  {Gasenzer}}]{Chantesana:2018qsb}%
  \BibitemOpen
  \bibfield  {author} {\bibinfo {author} {\bibfnamefont {I.}~\bibnamefont
  {Chantesana}}, \bibinfo {author} {\bibfnamefont {A.}~\bibnamefont
  {Piñeiro~Orioli}}, \ and\ \bibinfo {author} {\bibfnamefont {T.}~\bibnamefont
  {Gasenzer}},\ }\href {\doibase 10.1103/PhysRevA.99.043620} {\bibfield
  {journal} {\bibinfo  {journal} {Phys. Rev.}\ }\textbf {\bibinfo {volume}
  {A99}},\ \bibinfo {pages} {043620} (\bibinfo {year} {2019})}\BibitemShut
  {NoStop}%
\bibitem [{\citenamefont {Boguslavski}\ and\ \citenamefont {Pi\~neiro
  Orioli}(2020)}]{Boguslavski:2019ecc}%
  \BibitemOpen
  \bibfield  {author} {\bibinfo {author} {\bibfnamefont {K.}~\bibnamefont
  {Boguslavski}}\ and\ \bibinfo {author} {\bibfnamefont {A.}~\bibnamefont
  {Pi\~neiro Orioli}},\ }\href {\doibase 10.1103/PhysRevD.101.091902}
  {\bibfield  {journal} {\bibinfo  {journal} {Phys. Rev. D}\ }\textbf {\bibinfo
  {volume} {101}},\ \bibinfo {pages} {091902} (\bibinfo {year} {2020})},\
  \Eprint {http://arxiv.org/abs/1911.04506} {arXiv:1911.04506 [hep-ph]}
  \BibitemShut {NoStop}%
\bibitem [{\citenamefont {Schlichting}(2012)}]{Schlichting:2012es}%
  \BibitemOpen
  \bibfield  {author} {\bibinfo {author} {\bibfnamefont {S.}~\bibnamefont
  {Schlichting}},\ }\href {\doibase 10.1103/PhysRevD.86.065008} {\bibfield
  {journal} {\bibinfo  {journal} {Phys. Rev.}\ }\textbf {\bibinfo {volume}
  {D86}},\ \bibinfo {pages} {065008} (\bibinfo {year} {2012})}\BibitemShut
  {NoStop}%
\bibitem [{\citenamefont {Berges}\ \emph {et~al.}(2009)\citenamefont {Berges},
  \citenamefont {Scheffler},\ and\ \citenamefont {Sexty}}]{Berges:2008mr}%
  \BibitemOpen
  \bibfield  {author} {\bibinfo {author} {\bibfnamefont {J.}~\bibnamefont
  {Berges}}, \bibinfo {author} {\bibfnamefont {S.}~\bibnamefont {Scheffler}}, \
  and\ \bibinfo {author} {\bibfnamefont {D.}~\bibnamefont {Sexty}},\ }\href
  {\doibase 10.1016/j.physletb.2009.10.032} {\bibfield  {journal} {\bibinfo
  {journal} {Phys. Lett.}\ }\textbf {\bibinfo {volume} {B681}},\ \bibinfo
  {pages} {362} (\bibinfo {year} {2009})}\BibitemShut {NoStop}%
\bibitem [{\citenamefont {Kurkela}\ and\ \citenamefont
  {Moore}(2011)}]{Kurkela:2011ti}%
  \BibitemOpen
  \bibfield  {author} {\bibinfo {author} {\bibfnamefont {A.}~\bibnamefont
  {Kurkela}}\ and\ \bibinfo {author} {\bibfnamefont {G.~D.}\ \bibnamefont
  {Moore}},\ }\href {\doibase 10.1007/JHEP12(2011)044} {\bibfield  {journal}
  {\bibinfo  {journal} {JHEP}\ }\textbf {\bibinfo {volume} {12}},\ \bibinfo
  {pages} {044} (\bibinfo {year} {2011})}\BibitemShut {NoStop}%
\bibitem [{\citenamefont {Boguslavski}\ \emph {et~al.}(2019)\citenamefont
  {Boguslavski}, \citenamefont {Kurkela}, \citenamefont {Lappi},\ and\
  \citenamefont {Peuron}}]{Boguslavski:2019fsb}%
  \BibitemOpen
  \bibfield  {author} {\bibinfo {author} {\bibfnamefont {K.}~\bibnamefont
  {Boguslavski}}, \bibinfo {author} {\bibfnamefont {A.}~\bibnamefont
  {Kurkela}}, \bibinfo {author} {\bibfnamefont {T.}~\bibnamefont {Lappi}}, \
  and\ \bibinfo {author} {\bibfnamefont {J.}~\bibnamefont {Peuron}},\ }\href
  {\doibase 10.1103/PhysRevD.100.094022} {\bibfield  {journal} {\bibinfo
  {journal} {Phys. Rev. D}\ }\textbf {\bibinfo {volume} {100}},\ \bibinfo
  {pages} {094022} (\bibinfo {year} {2019})},\ \Eprint
  {http://arxiv.org/abs/1907.05892} {arXiv:1907.05892 [hep-ph]} \BibitemShut
  {NoStop}%
\bibitem [{\citenamefont {Moore}\ and\ \citenamefont
  {Turok}(1997)}]{Moore:1997cr}%
  \BibitemOpen
  \bibfield  {author} {\bibinfo {author} {\bibfnamefont {G.~D.}\ \bibnamefont
  {Moore}}\ and\ \bibinfo {author} {\bibfnamefont {N.}~\bibnamefont {Turok}},\
  }\href {\doibase 10.1103/PhysRevD.56.6533} {\bibfield  {journal} {\bibinfo
  {journal} {Phys. Rev. D}\ }\textbf {\bibinfo {volume} {56}},\ \bibinfo
  {pages} {6533} (\bibinfo {year} {1997})},\ \Eprint
  {http://arxiv.org/abs/hep-ph/9703266} {arXiv:hep-ph/9703266} \BibitemShut
  {NoStop}%
\bibitem [{Note1()}]{Note1}%
  \BibitemOpen
  \bibinfo {note} {We note that in classical-statistical simulations of highly
  occupied systems far from equilibrium the dependence on the ultraviolet modes
  can be further reduced if the characteristic dynamical scale $\sim Q_s$ is
  much lower.}\BibitemShut {Stop}%
\bibitem [{\citenamefont {Aarts}\ and\ \citenamefont
  {Berges}(2002)}]{Aarts:2001yn}%
  \BibitemOpen
  \bibfield  {author} {\bibinfo {author} {\bibfnamefont {G.}~\bibnamefont
  {Aarts}}\ and\ \bibinfo {author} {\bibfnamefont {J.}~\bibnamefont {Berges}},\
  }\href {\doibase 10.1103/PhysRevLett.88.041603} {\bibfield  {journal}
  {\bibinfo  {journal} {Phys. Rev. Lett.}\ }\textbf {\bibinfo {volume} {88}},\
  \bibinfo {pages} {041603} (\bibinfo {year} {2002})}\BibitemShut {NoStop}%
\bibitem [{\citenamefont {Smit}\ and\ \citenamefont
  {Tranberg}(2002)}]{Smit:2002yg}%
  \BibitemOpen
  \bibfield  {author} {\bibinfo {author} {\bibfnamefont {J.}~\bibnamefont
  {Smit}}\ and\ \bibinfo {author} {\bibfnamefont {A.}~\bibnamefont
  {Tranberg}},\ }\href {\doibase 10.1088/1126-6708/2002/12/020} {\bibfield
  {journal} {\bibinfo  {journal} {JHEP}\ }\textbf {\bibinfo {volume} {12}},\
  \bibinfo {pages} {020} (\bibinfo {year} {2002})},\ \Eprint
  {http://arxiv.org/abs/hep-ph/0211243} {arXiv:hep-ph/0211243} \BibitemShut
  {NoStop}%
\bibitem [{\citenamefont {Ambjorn}\ \emph {et~al.}(1991)\citenamefont
  {Ambjorn}, \citenamefont {Askgaard}, \citenamefont {Porter},\ and\
  \citenamefont {Shaposhnikov}}]{Ambjorn:1990pu}%
  \BibitemOpen
  \bibfield  {author} {\bibinfo {author} {\bibfnamefont {J.}~\bibnamefont
  {Ambjorn}}, \bibinfo {author} {\bibfnamefont {T.}~\bibnamefont {Askgaard}},
  \bibinfo {author} {\bibfnamefont {H.}~\bibnamefont {Porter}}, \ and\ \bibinfo
  {author} {\bibfnamefont {M.~E.}\ \bibnamefont {Shaposhnikov}},\ }\href
  {\doibase 10.1016/0550-3213(91)90341-T} {\bibfield  {journal} {\bibinfo
  {journal} {Nucl. Phys. B}\ }\textbf {\bibinfo {volume} {353}},\ \bibinfo
  {pages} {346} (\bibinfo {year} {1991})}\BibitemShut {NoStop}%
\bibitem [{\citenamefont {Boguslavski}\ \emph {et~al.}(2018)\citenamefont
  {Boguslavski}, \citenamefont {Kurkela}, \citenamefont {Lappi},\ and\
  \citenamefont {Peuron}}]{Boguslavski:2018beu}%
  \BibitemOpen
  \bibfield  {author} {\bibinfo {author} {\bibfnamefont {K.}~\bibnamefont
  {Boguslavski}}, \bibinfo {author} {\bibfnamefont {A.}~\bibnamefont
  {Kurkela}}, \bibinfo {author} {\bibfnamefont {T.}~\bibnamefont {Lappi}}, \
  and\ \bibinfo {author} {\bibfnamefont {J.}~\bibnamefont {Peuron}},\ }\href
  {\doibase 10.1103/PhysRevD.98.014006} {\bibfield  {journal} {\bibinfo
  {journal} {Phys. Rev.}\ }\textbf {\bibinfo {volume} {D98}},\ \bibinfo {pages}
  {014006} (\bibinfo {year} {2018})}\BibitemShut {NoStop}%
\bibitem [{Note2()}]{Note2}%
  \BibitemOpen
  \bibinfo {note} {More precisely, we obtain $\zeta =0.24\pm 0.03$ and $\zeta
  =0.25\pm 0.03$ by employing the method presented in \protect \Cref
  {app:zeta_estimate}.}\BibitemShut {Stop}%
\bibitem [{\citenamefont {Micha}\ and\ \citenamefont
  {Tkachev}(2003)}]{Micha:2002ey}%
  \BibitemOpen
  \bibfield  {author} {\bibinfo {author} {\bibfnamefont {R.}~\bibnamefont
  {Micha}}\ and\ \bibinfo {author} {\bibfnamefont {I.~I.}\ \bibnamefont
  {Tkachev}},\ }\href {\doibase 10.1103/PhysRevLett.90.121301} {\bibfield
  {journal} {\bibinfo  {journal} {Phys. Rev. Lett.}\ }\textbf {\bibinfo
  {volume} {90}},\ \bibinfo {pages} {121301} (\bibinfo {year} {2003})},\
  \Eprint {http://arxiv.org/abs/hep-ph/0210202} {arXiv:hep-ph/0210202}
  \BibitemShut {NoStop}%
\bibitem [{\citenamefont {Berges}\ \emph
  {et~al.}(2014{\natexlab{c}})\citenamefont {Berges}, \citenamefont
  {Boguslavski}, \citenamefont {Schlichting},\ and\ \citenamefont
  {Venugopalan}}]{Berges:2013lsa}%
  \BibitemOpen
  \bibfield  {author} {\bibinfo {author} {\bibfnamefont {J.}~\bibnamefont
  {Berges}}, \bibinfo {author} {\bibfnamefont {K.}~\bibnamefont {Boguslavski}},
  \bibinfo {author} {\bibfnamefont {S.}~\bibnamefont {Schlichting}}, \ and\
  \bibinfo {author} {\bibfnamefont {R.}~\bibnamefont {Venugopalan}},\ }\href
  {\doibase 10.1007/JHEP05(2014)054} {\bibfield  {journal} {\bibinfo  {journal}
  {JHEP}\ }\textbf {\bibinfo {volume} {05}},\ \bibinfo {pages} {054} (\bibinfo
  {year} {2014}{\natexlab{c}})},\ \Eprint {http://arxiv.org/abs/1312.5216}
  {arXiv:1312.5216 [hep-ph]} \BibitemShut {NoStop}%
\bibitem [{Note3()}]{Note3}%
  \BibitemOpen
  \bibinfo {note} {We have checked that other powers like $\protect \frac
  {1}{(\gamma L)^2} \DOTSI \intop \ilimits@ _0^{\gamma L} d\Delta x\protect
  \,\Delta x$ lead to similar scaling exponents.}\BibitemShut {Stop}%
\bibitem [{\citenamefont {Schmied}\ \emph {et~al.}(2019)\citenamefont
  {Schmied}, \citenamefont {Mikheev},\ and\ \citenamefont
  {Gasenzer}}]{Schmied:2018mte}%
  \BibitemOpen
  \bibfield  {author} {\bibinfo {author} {\bibfnamefont {C.-M.}\ \bibnamefont
  {Schmied}}, \bibinfo {author} {\bibfnamefont {A.~N.}\ \bibnamefont
  {Mikheev}}, \ and\ \bibinfo {author} {\bibfnamefont {T.}~\bibnamefont
  {Gasenzer}},\ }\href {\doibase 10.1142/S0217751X19410069} {\bibfield
  {journal} {\bibinfo  {journal} {Int. J. Mod. Phys. A}\ }\textbf {\bibinfo
  {volume} {34}},\ \bibinfo {pages} {1941006} (\bibinfo {year} {2019})},\
  \Eprint {http://arxiv.org/abs/1810.08143} {arXiv:1810.08143
  [cond-mat.quant-gas]} \BibitemShut {NoStop}%
\bibitem [{\citenamefont {Mikheev}\ \emph {et~al.}(2019)\citenamefont
  {Mikheev}, \citenamefont {Schmied},\ and\ \citenamefont
  {Gasenzer}}]{Mikheev:2018adp}%
  \BibitemOpen
  \bibfield  {author} {\bibinfo {author} {\bibfnamefont {A.~N.}\ \bibnamefont
  {Mikheev}}, \bibinfo {author} {\bibfnamefont {C.-M.}\ \bibnamefont
  {Schmied}}, \ and\ \bibinfo {author} {\bibfnamefont {T.}~\bibnamefont
  {Gasenzer}},\ }\href {\doibase 10.1103/PhysRevA.99.063622} {\bibfield
  {journal} {\bibinfo  {journal} {Phys. Rev.}\ }\textbf {\bibinfo {volume}
  {A99}},\ \bibinfo {pages} {063622} (\bibinfo {year} {2019})}\BibitemShut
  {NoStop}%
\end{thebibliography}%

\end{document}